\documentclass[12pt,preprint]{aastex}
\usepackage{lscape}
\shorttitle{VSOP Survey Images}
\shortauthors{Scott et al.}

\begin{document}
\title{THE VSOP 5~GHz AGN SURVEY: \\
III. IMAGING RESULTS FOR THE FIRST 102 SOURCES}

\author{W. K. Scott           \altaffilmark{1},
        E. B. Fomalont        \altaffilmark{2},
        S. Horiuchi           \altaffilmark{3,4,5},
        J. E. J. Lovell       \altaffilmark{6},
        G. A. Moellenbrock    \altaffilmark{7},
        R. G. Dodson          \altaffilmark{8,9},
        P. G. Edwards         \altaffilmark{9},
        G. V. Coldwell        \altaffilmark{10},
        S. Fodor              \altaffilmark{4,11},
        S. Frey               \altaffilmark{12},
        L. I. Gurvits         \altaffilmark{13},
        H. Hirabayashi        \altaffilmark{9},
        M. L. Lister          \altaffilmark{2},
        L. Mosoni             \altaffilmark{14,12},
        Y. Murata             \altaffilmark{9},
        Z. Paragi             \altaffilmark{12,13},
        B. G. Piner           \altaffilmark{11},
        Z.-Q. Shen            \altaffilmark{15,9,16},
        A. R. Taylor          \altaffilmark{1},
        S. J. Tingay          \altaffilmark{5,6}}

\altaffiltext{1}{Physics and Astronomy Department, University of Calgary,
                 2500 University Dr. NW,
                 Calgary, Alberta, Canada, T2N 1N4 
                 (bill@ras.ucalgary.ca, russ@ras.ucalgary.ca)}

\altaffiltext{2}{National Radio Astronomy Observatory, 520 Edgemont Road,
                 Charlottesville, VA 22903, USA  
                 (efomalon@nrao.edu, mlister@nrao.edu)}

\altaffiltext{3}{National Astronomical Observatory of Japan, 2-21-1 Osawa, 
                 Mitaka, Tokyo 181-8588, Japan}

\altaffiltext{4}{Jet Propulsion Laboratory, 4800 Oak Grove Drive,
                 Pasadena, CA 91109, USA}

\altaffiltext{5}{Centre for Astrophysics and Supercomputing,
                  Swinburne University of Technology,
                  P.O. Box 218, Hawthorn, VIC 3122, Australia
                  (stingay@astro.swin.edu.au, shoriuchi@astro.swin.edu.au)}

\altaffiltext{6}{Australia Telescope National Facility,
                 Commonwealth Scientific and Industrial Research Organization,
                 P. O. Box 76, Epping NSW 2122, Australia
                 (Jim.Lovell@csiro.au)}

\altaffiltext{7}{National Radio Astronomy Observatory, 
                 P.O. Box 0, Socorro, NM 87801, USA
                 (gmoellen@nrao.edu)}

\altaffiltext{8}{School of Mathematics and Physics, 
                 University of Tasmania, GPO Box 252-21,
                 Hobart, Tasmania 7001, Australia}

\altaffiltext{9}{Institute of Space and Astronautical Science, 
                 Japan Aerospace Exploration Agency,
                 3-1-1 Yoshinodai,
                 Sagamihara, Kanagawa 229-8510, Japan
                 (rdodson@vsop.isas.jaxa.jp, pge@vsop.isas.jaxa.jp,
                  hirax@vsop.isas.jaxa.jp,  murata@vsop.isas.jaxa.jp) }

\altaffiltext{10}{Grupo de Investigaciones en Astronomia Te\'{o}rica y Experimental, 
                 Observatorio Astron\'{o}mico, 
                 Universidad Nacional de C\'{o}rdoba,
                 Laprida 854, 5000, C\'{o}rdoba, Argentina
                 (georgina@oac.uncor.edu) }

\altaffiltext{11}{Physics Department, Whittier College,
                  13406 East Philadelphia, P.O. Box 634, Whittier,
                  CA 90608-4413, USA
                  (gpiner@whittier.edu, fodor\_zie@yahoo.com)} 

\altaffiltext{12}{F\"{O}MI Satellite Geodetic Observatory, P.O. Box 585,
                 H-1592, Budapest, Hungary
                 (frey@sgo.fomi.hu)}

\altaffiltext{13}{Joint Institute for VLBI in Europe, P.O. Box 2,
                 7990 AA, Dwingeloo, The Netherlands
                 (lgurvits@jive.nl, zparagi@jive.nl)} 

\altaffiltext{14}{MTA Konkoly Observatory,
                 P.O. Box 67, H-1525, Budapest, Hungary
                 (mosoni@konkoly.hu)} 

\altaffiltext{15}{Shanghai Astronomical Observatory, National Astronomical
                  Observatories, Chinese Academy of Sciences,
                  Shanghai 200030, China
                  (zshen@center.shao.ac.cn)}

\altaffiltext{16}{Institute of Astronomy and Astrophysics, Academia Sinica,
                  P.O. Box 23-141
                  Taipei 106, Taiwan, China}
\begin{abstract}

The VSOP mission is a Japanese-led project to study radio sources with
sub-milliarcsec resolution using an orbiting 8 m telescope, HALCA,
along with global arrays of Earth-based telescopes. 
Approximately 25\% of the observing
time is devoted to a survey of compact AGN
which are stronger than 1 Jy at 5~GHz---the VSOP AGN Survey.
This paper, the
third in the series, presents the results from the analysis of the
first 102 Survey sources.
We present high resolution
images and plots of visibility
amplitude versus projected baseline length. In addition, 
model-fit parameters to the primary radio components are listed,
and from these 
the angular size and 
brightness temperature for the radio cores are calculated.
For those sources for which we were able to determine the
source frame core brightness temperature,
a significant fraction (53 out of 98) have a source frame core brightness
temperature in excess of $10^{12}$\,K.
The maximum source frame core brightness temperature we observed
was $1.2\times10^{13}$~K. Explaining a brightness temperature this high
requires an extreme amount of 
relativistic Doppler beaming.
Since the maximum brightness temperature one is able to determine
using only ground-based arrays is of the order of $10^{12}$~K,
our results confirm the necessity of using space VLBI
to explore the extremely high brightness temperature regime.

\end{abstract}
\section{INTRODUCTION}

On 1997 February 12, the Institute of Space and Astronautical Science
launched the HALCA (Highly Advanced Laboratory for Communications and
Astronomy) satellite, which contained
an 8 m telescope, dedicated specifically
to VLBI.  With an apogee height of 21400 km, radio sources are imaged
with angular resolution three times greater than with ground arrays at
the same frequency \citep{hir98}. About 25\% of the observing time
was dedicated to the VSOP (VLBI Space Observatory Programme)
Survey of approximately 400 flat-spectrum
AGN which are stronger than 1 Jy at 5 GHz.
Those 294 sources which had compact structures suitable for
observations with VSOP
were designated the VSOP Source Sample (VSS) \citep{fom00a, edw02}.
The compilation
and general description of the VSOP AGN Survey is given by
\cite{hir00b} (Paper I) and \cite{fom00a}.  The major goal of the
Survey is to determine statistical properties of the 
sub-milliarcsecond
structure of the strongest extragalactic radio sources at 5 GHz, and
to compare these structures with other properties of the sources.
Combined with ground observations at many radio frequencies
(single-dish and VLBI), and at higher energies, the Survey will provide
an invaluable list for further detailed ground-based studies, as well
as a list of sources to be observed in future space VLBI missions.

This paper is the third of the VSOP Survey series.
Paper II \citep{lov04} describes the reduction of the data and the
specific problems associated with reducing space VLBI data.
In this paper, we present the
results of 102 sources in the VSS list for which the data reduction
is now complete. 
The results include an image and a simple model for each source, and the
observed distribution of radio core sizes and brightness temperatures.
Paper IV
\citep{hor04} contains a statistical analysis using the visibility
data.

In \S 2 we briefly describe the source list compilation, the
observations, and data reduction. 
This information is provided in more detail in other VSOP
Survey papers.  The presentation of the results is given in graphical
and tabular form in \S 3.  The discussion of the source structures,
with emphasis on the radio cores and source brightness
temperatures, is given in \S 4.

\section {SOURCE SAMPLE, OBSERVATIONS, AND DATA REDUCTION}
 
The description of the VSOP mission and the 5 GHz AGN Survey has
been given in \cite{hir98,fom00b,hir00a,hir00b}, and
therefore will only be briefly summarized here.  In order to be
included in the VSOP Survey a source
was required to have:
\newline $\bullet$ a total flux density at 5 GHz, $S_5\geq
5.0$~Jy\newline \centerline{or\phantom{aaaaaaaaaaaaaaaaaaaaaaa}}
$\bullet$ a total flux density at 5 GHz, $S_5\geq 0.95$~Jy and\newline
$\bullet$ a spectral index $\alpha \geq -0.45$ ($S \propto
\nu^\alpha$) and\newline $\bullet$ a galactic latitude $|b|\geq
10^\circ$.

The finding surveys from which sources were selected were primarily the 
Green Bank GB6 Catalog for the
northern sky \citep{gre96}, and the Parkes-MIT-NRAO (PMN) Survey
\citep{law86,gri93} for the southern sky. 
The 402 sources satisfying these criteria comprise the VSOP source list
\citep{hir00b}.

However, it was expected that 
many sources in the VSOP source list would not be detectable by
HALCA due to insufficient correlated flux on the longest baselines. Hence, most
sources with declination $>-44^\circ$ were 
observed in a VLBA pre-launch survey (VLBApls)
\citep{fom00b}.
Based upon the VLBApls results
a cutoff criterion, a minimum flux density of $0.32$\,Jy
at 140~M$\lambda$, was established for inclusion of a source in the
Survey \citep{fom00a}.
Those sources exceeding this threshold, as well as those sources
south of  $-44^\circ$, were designated the VSOP
Source Sample (VSS), and were scheduled for VSOP observations.
The VSS contains 294 sources, 289 in the original sample
\citep{hir00b}, plus 5 extra sources which were added after it was
discovered that their VLBA observations suffered from 
insufficiently accurate positional information \citep{edw02}.
These sources do not comprise a
rigidly defined complete sample, and the VSS
is discussed in more detail in Paper IV where a statistical analysis
of the visibility data is undertaken.

Observations of the VSS began in August 1997,
with the most recent observations being made
in October 2003. 
However, a subsequent
loss of attitude control of the HALCA satellite
has prevented any further observations as of the time of writing
(although it is hoped that attitude control may again
be regained, in which case Survey observations will
be resumed).
In this paper we present the analysis of the 102
experiments successfully reduced by the end of 2002.
The
observing parameters for these 102 experiments can be found in Table 1.
A typical experiment uses two to five ground-based telescopes plus
HALCA, with a total observation time per source
of approximately four hours. 
In addition, for
those sources in the VSS which were also included in
scientific general observing time (GOT) proposals,
a subset of the data was extracted
for use in the Survey. To be
comparable with the Survey data were extracted from typically three or
four GRTs (ground radio telescopes), 
and covered about four hours of observation time.
Approximately 100 sources in the VSS have GOT extracted data.
Of the 102 sources described in this paper, 56 have GOT extracted data.
The GRTs of choice for most data extraction experiments 
with the VLBA as the ground-based array were
Mauna Kea and St. Croix, the GRTs which provided the longest ground
baselines.

A chart indicating the status of all the 289 Survey experiments as of
the end of 2002 is given in Fig.\ 1. A somewhat higher percentage of
stronger Survey sources are completed because the data for many of
these were extracted from GOT
proposals, and these had higher priority in scheduling than Survey
experiments. Typically, these experiments tended to look at the stronger
\lq\lq famous" radio sources.
There was also some attempt to schedule with a higher priority those
Survey sources
having a total flux density greater than 1.3 Jy. 
Nevertheless, the date for
scheduling any specific experiment was randomized to some extent.
Firstly, there were HALCA constraints which limited the area of sky that
could be observed on any given date. 
Secondly, the period when reasonable $(u,v)$
coverage for the target source could be achieved had an 18-month
cycle.  Thus, the distribution of the 102 sources given in this paper
should be reasonably representative of the AGN sample.

All calibration and editing of the data were carried out using the AIPS software
package \citep{gre88}, while Difmap \citep{she97} was used to image the data.
The calibration, editing, and imaging of the VSOP Survey data is
described fully elsewhere \citep{moe00,lov04}.  Since most of the
sources, especially those with $\delta > -44^\circ$,
have been imaged with previous ground VLBI observations,
consistency of the VSOP image with these other images was used to
constrain the cleaning and modeling. The results of, and supporting
documentation for the data reduction can be found on the web site
(http://www.vsop.isas.jaxa.jp/survey). The calibrated data are
available from ISAS on request.
\section{THE RESULTS}

\subsection {The Visibility Amplitudes and Images}

The graphical results of the data 
reduction are shown in Fig.\ 2. For each source
three separate panels are presented horizontally across the page,
the $(u,v)$ coverage, the correlated flux density
versus $(u,v)$ radius, and the cleaned image.
The
quality of the images varies considerably. 
However, even for sources which have
only two GRTs, the major structural details of the core
can be ascertained, albeit with some loss
of sensitivity to larger-scale radio emission.

The fidelity
of these images is limited by two factors, the effect of the $(u,v)$
plane undersampling, and the uncertainty in the amplitude
calibration.

Because of the 
length of time necessary for the satellite to slew between sources,
HALCA was not able to participate in fringe-finder or flux-calibrator
scans scheduled in the VSOP observations.
Amplitude self-calibration was possible for images
made with the data from four or more antennas. 
However, for the Survey the number of baselines was often
insufficient to perform this. The amplitude calibration is therefore
often entirely derived from the measured or expected 
gain and system temperatures of
the antennas.

Although the amplitude scale of the VSOP Survey data was calibrated
during the data reduction stage \citep{lov04}, 
the visibility amplitudes of
the 102 sources in this paper were compared with those found in the
VLBApls in order to find any remaining systematic
flux density offsets.
We compared the flux densities we found with those from the 
VLBApls at
nearby points on the $(u,v)$ plane (less than 10~M$\lambda$
apart). This allows sources with significant structure to be compared
directly. To take account of the high variability often seen
in these sources we corrected for the change in flux density
between the VLBApls and the Survey experiments
by using data from 
the Australia Telescope Compact Array (ATCA) 
\citep{tin03}
or the University of Michigan Radio Astronomy Observatory 
(UMRAO)\footnote{http://www.astro.lsa.umich.edu/obs/radiotel/umrao.html}
monitoring programs.
This variability information was available for approximately 80\%
of the Survey sources.
This comparison is at best a zeroth
order correction, as it can only correct for the flux density changes in the
object as a whole, while the 
individual $(u,v)$ sample points will be affected
differently by the variability.

We selected those experiments for which we had some measure of the
source variability. In addition, to exclude the most extremely variable sources,
we required the VLBApls and Survey visibilities to agree to within
a factor of 2.5.
Finally, we restricted ourselves to only comparing visibility points for
which the 
correlated flux density on baselines to HALCA
was greater than 0.5~Jy. 

We found that the median ratio
of the Survey to VLBApls correlated flux 
density 
was $0.83$ with an error in the median of approximately $0.05$.
The origin of this
discrepancy between the two surveys is not understood at this time.
All the VSOP Survey visibilities have 
been rescaled upwards by $1.2$, the reciprocal
of this factor.
The median ratio for VLBA data extraction experiments was slightly 
closer to unity than for non-VLBA experiments.
As the system temperatures and gain curves for the VLBA
GRTs are significantly better than those from non dedicated VLBI
arrays this is of no surprise.
Comparing the VLBApls to ATCA monitoring data 
\citep{tin03}, and the UMRAO database,
we find
that the median ratio is close to unity (1.03 with an error in
the median of 0.04).

The second factor affecting the image fidelity is the 
uneven and often poor sampling of the $(u,v)$ plane.
The effect of uneven $(u,v)$ sampling has been investigated by
\cite{lis01}. In simulations involving $\rm{VLBA}+\rm{HALCA}$
$(u,v)$ plane coverage they found 
that the peak value in their simulated images divided by the maximum
difference between the simulated images and the original model
was between approximately 30:1 to 100:1.
This is probably a much better indicator of the image fidelity than
the dynamic range, which was much higher for their images ($\ge$ 1000:1).
For our images the dynamic range and image fidelity will almost certainly
be less than their values, due to the larger uncertainty often present in the
relative visibility amplitude scaling, and the smaller number of baselines.

Hence, based upon the findings of \citet{lis01}, 
and after applying the overall systematic error factor of $20\%$,
we conservatively estimate that
our image fidelity is in the 20:1 range, i.e.
caution should be exercised when interpreting any features
in the images at approximately the 
$5\%$ level of the peak flux density, or less.
For the images constructed from GOT data using the VLBA this figure may
be higher, while for images from Survey observations using non-VLBA 
antennas this figure may be as low as 10:1, due to additional uncertainty
in the amplitude scaling of the GRTs.

This amplitude scaling uncertainty will also affect the peak flux densities
listed for the images. The $1\sigma$ scatter in the 
ratios
of the Survey to VLBApls correlated flux
density was approximately 0.2. A portion of this scatter will be
due to source variability, for which we only made a single, global
correction for each experiment. However, if we assume that all the scatter
is simply due to errors in the visibility amplitudes, then this gives
us a conservative error estimate. The errors in the peak flux values for
each image will be comparable, ie. approximately 20\%.

\subsection{Model-Fitting and Brightness Temperature Determination}

After obtaining the best image of a source
the visibilities were model-fit with Difmap to a small number of
components, typically a combination of two or three $\delta$-functions,
circular Gaussians, or elliptical Gaussians. The precise number and type
of components were chosen to minimize the $\chi^2$, while at the same time
keeping the number of free parameters as small as possible.
Minimizing the number of components 
(ie. model-fitting to only the strongest components)
also helps to reduce the possibility
of model-fitting a component which is not, in fact, real. 
In almost all
cases the integrated flux of any weaker components in a source was greater
than $5\%$ of the integrated flux of the strongest component.
The only exceptions were for sources J1229$+$0203 and J1407$+$2827. In 
both of these cases a weak, point source component was necessary
to fit the visibilities on the longer baselines.

In order to ensure that the lower sensitivity HALCA data were not
largely ignored during the model-fitting they were up-weighted
to give them a weight comparable to
the ground baseline data \citep{lov04}.
The parameters for all the components are listed in Table 2.
The flux density errors due to the fitting
are typically 0.03 Jy, or 10\% of the
component flux density, whichever is greater, giving a total
error in the flux density parameters (including the error
in the overall visibility amplitude scaling) of
approximately 0.03 Jy, or 25\% of the component flux density.

The lower error bound
for the diameter of a component is given by:
\begin{equation} 
D_{-}=\Bigl[[D^2 + B^2][1-(1/S)]^2-B^2\Bigr]^{1/2}
\end{equation} 
where $D$ is the FWHM diameter of the component,
$B$ is HPBW in the direction of $D$,
and $S$ is the signal to noise ratio associated with the component
\citep{fom00b}.
When the argument of the square-root is negative, 
then the lower error bound on the diameter
is zero. In this case, the component was assumed to be consistent
with a zero diameter component (either a $\delta$-function or 
zero axial ratio elliptical Gaussian), and the diameter was set to zero.


For each source the
brightness temperature of the core was
determined using the model-fit components.
The brightness temperature of a component in the observer's frame
is given by:
\begin{equation}
T_b=\frac{S\lambda^2}{2 k_b \Omega},
\label{eqtb}
\end{equation}
where $S$ is the component flux density 
at wavelength $\lambda$, $k_b$ is Boltzmann's
constant, and $\Omega\approx 1.13 \cdot\rm{(B_{maj})(B_{min})}$
is the solid angle subtended by the component
(which we have expressed in terms of the full width at half maximum of the
component major and minor axes).
To convert to brightness temperature in the source frame, equation
\ref{eqtb} is multiplied by $(1+z)$, where $z$ is the source redshift.

The error in the $\rm{B_{maj}}$ and $\rm{B_{min}}$ component parameters,
when the core is resolved,
is approximately 10\%, giving
an error in the solid angle subtended by the components of
approximately 15\%. When combined with the approximately 25\%
error in the component flux densities, this implies an error in
the brightness temperature
for resolved components
of about 30\%.

The determination of the core component was
based upon the morphology
of the source, VLBI images of the source at other frequencies, and/or
identifications from external references.
With the exception of J1723$-$6500, a GPS source whose core was
probably not detected,
the core brightness temperatures are
listed in Table 2.
However, considerable caution
should be exercised when the core component overlaps one or more other
components. Overlapping components are denoted by a \lq\lq $*$" or 
\lq\lq $\dag$" in the flux density column in Table 2. We have
defined \lq\lq overlapping" to mean that a component's position
(in the case of a $\delta$-function) or FWHM is located within,
or overlaps the FWHM of another component.
In addition, caution should be exercised in any interpretation of the
major axis position angle of the core component. 
\cite{lis01} has found that 
for unequal $(u,v)$ plane coverage, the
major axis position angle of the core component
is strongly correlated with that of the beam, and
is therefore unlikely to reflect an intrinsic property of the source.

A lower bound on the brightness temperature
of unresolved cores
was determined using the Difwrap software package \citep{lov00b}.
Difwrap is a Perl language graphical user interface for the Difmap
software package.
Using Difwrap, the (component flux density, component size)
parameter space was searched to find the combination of
parameters which had the
lowest
brightness temperature, but still had an adequate fit to the data (based
upon plots of visibility amplitude versus $(u,v)$ radius).
For a fit to be adequate the model still had to
retain the essential structure of the amplitude versus $(u,v)$
plot. However, small deviations between the model and the visibility
data, of the order of 20\% or so, were allowed.
For each point in the parameter space the best fit in position
was determined using the Difmap {\tt modelfit} command.
All other components were kept fixed.
Although subjective, visually comparing the model-fits to the data
is thought to be a conservative estimate of the parameter errors
\citep{tzi89}.

The error estimate for the brightness temperature lower limits
will be greater than the error in the brightness temperature
of resolved sources, due to the additional subjectivity
in assessing what comprises an \lq\lq adequate fit".
An error of $\pm 50\%$ is probably reasonable.

Histograms depicting the frequency of occurrence of the brightness
temperature in both the source and observer's frame are shown in
Fig.\ 3. 
Most cores have $T_b>10^{11}$~K, 
with approximately 54\% of the sources having a
brightness temperature (or brightness temperature lower limit)
in excess of $10^{12}$~K in the source frame, 
and approximately 38\% of the
sources having a brightness temperature 
(or brightness temperature lower limit)
greater than $10^{12}$~K in the observer's frame.
The source with the highest core brightness temperature in the
source frame is J0539$-$2839, with $T_b= 1.2\times10^{13}$ K.

We can compare our brightness temperature distribution with
distributions obtained by other groups using space VLBI data.
The first brightness temperature determinations using space VLBI
were calculated from observations
using a 4.9~m diameter orbiting antenna, which was part of
the Tracking and Data Relay Satellite System (TDRSS) \cite{lev89}.
At 2.3~GHz, \citet{lin89} found that ten out of 14 sources (71\%) 
had source frame brightness temperatures in excess of $10^{12}$~K,
while at 15~GHz six out of nine sources (66\%) had source frame
brightness temperatures in excess of $10^{12}$~K \citep{lin90}.
Considering the relatively small number of sources in these samples,
this is in reasonable agreement with our results.
In addition, using general observing time VSOP data,
brightness temperatures
have been calculated
for a selection
of sources from the Pearson-Readhead source sample \citep{pea88}.
\citet{lis01} has calculated the observer-frame brightness temperature
for a selection of 27 of these sources, while \citep{tin01} has calculated
the comoving frame brightness temperature for 31 sources from this sample.
Approximately 26\% of the source frame brightness
temperatures were
in excess of $10^{12}$~K, while
approximately  19\% of the sources had comoving frame brightness temperatures
in excess of $10^{12}$~K.
While this latter result is considerably 
smaller than our result, \citet{tin01} normalized all their brightness
temperatures to brightness temperatures at a fixed frequency in the
comoving frame, which reduced their brightness temperatures by a factor
of $(1+z)^{1/2}$ when compared to our results.
In addition, \citet{tin01} calculated source brightness temperatures
assuming an optically thick core, which reduced their brightness temperatures
by an additional factor of $0.56$ with respect to our results.
Multiplying the \citet{tin01} brightness temperatures by 
$(1+z)^{1/2}/0.56$ 
results in about 48\% of their source frame brightness temperatures
exceeding $10^{12}$~K, approximately the same percentage as was seen in
this paper.

The distribution of the angular size
of the cores is plotted in Fig.\ 4.
The median angular size subtended by 
the resolved cores is about $0.075~\rm{mas}^2$,
corresponding to an angular diameter (FWHM) of approximately 0.26~mas.
Of course, the former value only represents a median of core
sizes for which the interferometer was sensitive to. 
Any cores with extremely small angular size ($\rm{FWHM}\approx 0.1~\rm{mas}$ or less, for instance)
are unresolved, and have not been incorporated into the calculation
of the median. In addition, any large-scale structure around the core 
may not have been detected (depending upon the 
$(u,v)$ plane coverage on shortest baselines), which may have rendered
the core size estimation for several of the sources artificially low.
Hence, any physical interpretation of the median core size should be treated
with caution.

\subsection{Comments on Individual Sources}

   When applicable, short notes on the sources are given. 
In addition, a general comparison is made with VLBI
images from other sources,
primarily the VLBApls \citep{fom00b},
U. S. Naval Observatory Database (USNO) \citep{fey96, fey97, fey00},
a space VLBI Survey of Pearson-Readhead sources (VSOPPR) \citep{lis01},
the VLBA 2~cm Survey (VLBA2cm1) \citep{kel98, zen02} and (VLBA2cm2) 
(L. I. Gurvits, K. I. Kellermann, E. B. Fomalont, \& H. Y. Zhang, 
in preparation), and results from VSOP observations 
of southern sources (VSOPsth) \citep{tin02}. Any
significant differences are noted.
When the core is unresolved, a limit on the area
subtended by
the core
is given.


\smallskip\noindent {\bf J0006$-$0623:}
The data quality for this source is relatively poor.
The extended emission is only fit to a single component due
to a paucity of baselines less than 100~M$\lambda$ in length.
The area subtended by the core is $<0.2~\rm{mas}^2 $. 

\smallskip\noindent {\bf J0019$+$7327:}
An image with the full GOT data-set is given in VSOPPR.
The VLBA2cm1 images show significant 
evolution of this source between
1994 and 2000.  The size and orientation of the VSOP image is
consistent with the VLBA2cm1 image in March 1998.
The area subtended by the core component is $<0.03~\rm{mas}^2 $.

\smallskip\noindent {\bf J0042$+$2320:} The VSOP image contains
much less flux density than that in the VLBApls and USNO. It may
be variable. 
The USNO 8.6~GHz image shows an extension 3~mas to
the north-northeast, and an elongation of the core to the south.
The VLBA2cm2 image has an additional component to the south.
The VLBApls image, however, shows an extension to the east.

\smallskip\noindent {\bf J0106$-$4034:}
The area subtended by the core component is $<0.05~\rm{mas}^2 $.
 
\smallskip\noindent {\bf J0115$-$0127:} 
The central region has been resolved into two separate components.
We have assumed the more northern component is the core.
The PA of the jet in the Survey image is slightly more
northerly than the PA of the jet 
seen in lower resolution images.

\smallskip\noindent {\bf J0121$+$1149:}
In 5~GHz observations by \citet{gab99} the core was unresolved,
although in the VLBA2cm1 and USNO images 
the core has an asymmetry towards the north. In addition, there is extended
emission north of the core in the VLBApls and USNO 2~GHz images.

\smallskip\noindent {\bf J0126$+$2559:} The data quality is poor, but
the VSOP image is consistent with those of
the VLBApls and USNO images.
The area subtended by the core component is $<1~\rm{mas}^2 $.

\smallskip\noindent {\bf J0136$+$4751:} 
Most of the emission is in a small, central component.  The emission to the
north is strong at low frequencies, and the component to the
north-west is stronger at high frequencies.  The VLBA2cm1 observations
show this north-west component decreasing in flux density with time.

\smallskip\noindent {\bf J0210$-$5101:} 
An image with the full GOT data-set
is given in VSOPsth. We have assumed the more compact northeastern
component is the core. 

\smallskip\noindent {\bf J0217$+$7349:}
The central region is resolved into two closely spaced components. We have
assumed the core is the more compact of the components.
An image made using all the GOT data can be found in VSOPPR.

\smallskip\noindent {\bf J0251$+$4315:} 
The USNO, VLBApls, and VLBA2cm2 show a large component around 15~mas
to the south-east.
We have assumed that the core is the most northwesterly component of
the strong, central emission region.

\smallskip\noindent {\bf J0319$+$4130:} (3C84) Images using
GOT data can be found in VSOPPR and \cite{asa00}. The
six component model fit does not account for some 
of the large-scale structure.
Following \citet{ver94} we have identified the core as the
most northeasterly component of the strong, central emission region.

\smallskip\noindent {\bf J0334$-$4008:} 
The redshift for this source is quoted 
from \cite{hew87}, although there appears to be little
support for this value. \cite{dri97} were unable to find a redshift
for this source.
The area subtended by the core component is $<0.07~\rm{mas}^2 $.


\smallskip\noindent {\bf J0403$-$3605:} The core is extended to the
north-east, in the direction of faint extended emission seen by the
VLBApls, USNO, and VLBA2cm2 images.

\smallskip\noindent {\bf J0405$-$1308:} 
Extended emission to the south, as seen in the VLBApls 
and VLBA2cm2 images, is not
seen in the VSOP image, possibly due to the lack of 
VSOP data on short spacings. 

\smallskip\noindent {\bf J0423$-$0120:} 
In the USNO and VLBA2cm1 images there is additional emission
south of the
core component. 

\smallskip\noindent {\bf J0440$-$4333:} An image made from the entire
GOT data-set is found in VSOPsth.
We have separated the emission into two components,
one of which is unresolved.  The one-component fit used in VSOPsth
contains both components reported here.

\smallskip\noindent {\bf J0453$-$2807:}
The inner jet points in the direction of the larger-scale structure
seen in the VLBApls.

\smallskip\noindent {\bf J0457$-$2324:} 
VLBApls image has a faint component extending off the core to the west.

\smallskip\noindent {\bf J0501$-$0149:} We have assumed the core is the
stronger component to the southeast.

\smallskip\noindent {\bf J0538$-$4405:} An image made from all the GOT
data is found in VSOPsth.

\smallskip\noindent {\bf J0539$-$2839:} The core is resolved and is
elongated toward the extended structure seen in the VLBApls.
ATCA light curves \citep{tin03}
indicate a dramatic increase in flux 
over the 40 months
between the VLBApls
and the VSOP observation.
The VLBA2cm2 image has a small additional component to the east.

\smallskip\noindent {\bf J0542$+$4951:} (3C147)  The
source is composed of a core component with diffuse emission to the
west. See also the VSOP image from \cite{sly01}.
The area subtended by the core is $<0.10~\rm{mas}^2 $

\smallskip\noindent {\bf J0555$+$3948:}
The jet to the west,
seen in the VLBA2cm1 and VLBA2cm2 images,
is clearly seen in the VSOP image. 
The data were very well modeled by two circular Gaussians,
one each for the core and jet components.


\smallskip\noindent {\bf J0609$-$1542:}
The core, as seen in the VLBApls image, is now clearly resolved out into
two distinct components.
Emission in the core region, as seen in the VSOP
image as well as higher frequency USNO ground images, is mostly in an
easterly direction. Lower resolution images show the jet taking a more
northeasterly tack farther from the core.
The area subtended by the core is $<0.05~\rm{mas}^2 $

\smallskip\noindent {\bf J0635$-$7516:}
An image made from all the GOT data is found in
VSOPsth. The core is complex and extended. We have used the easternmost
component in our determination of core brightness temperature.



\smallskip\noindent {\bf J0741$+$3112:} 
This is a GPS source \citep{sta97}.
The VLBApls image is resolved out into two components in our Survey image.
However, the PA of the eastern extension 
of the southern component is not consistent
with PA of a similar extension in the VLBApls, USNO, and VLBA2cm1 images,
probably due to insufficient short spacing data.
The southern component is the most compact, hence,
this may be the core.
The redshift and optical ID is taken from \cite{hew93}.

\smallskip\noindent {\bf J0748$+$2400:}
The area subtended by the core is $<0.2~\rm{mas}^2 $.

\smallskip\noindent {\bf J0811$+$0146:}
USNO images indicate a possible bent jet morphology, with the jet
direction towards the south near the core, 
then bending towards the west. The VSOP image of the core region is consistent
with this morphology.

\smallskip\noindent {\bf J0818$+$4222:} An image made from all the GOT
data is found in
VSOPPR. 
The redshift was obtained from \cite{law96}.

\smallskip\noindent {\bf J0824$+$5552:} 
Both 2 and 8~GHz USNO images show a jet aligned in a northeasterly
direction.


\smallskip\noindent {\bf J0841$+$7053:} 
An image made from the entire GOT data-set can be found in
VSOPPR. This is a complex source with an faint unresolved core.
We have identified the core to be the most northerly component.
The redshift was obtained from \cite{jac02}.
The area subtended by the core is $<0.1~\rm{mas}^2 $.

\smallskip\noindent {\bf J0854$+$2006:} (OJ287) 
A VSOP GOT image is given in
\cite{gab01}, who identify the
core as
the unresolved component
to the northeast of the strongest component.
The area subtended by the core is $<0.3~\rm{mas}^2 $.

\smallskip\noindent {\bf J0903$+$4651:}
The jet to the north seen in other VLBI images is not seen in our VSOP
image, probably due to the poor $(u,v)$ plane coverage. The complex core
region has been resolved into two closely spaced components.

\smallskip\noindent {\bf J0909$+$0121:} There is
some emission seen toward the larger-scale structure seen in the 
VLBApls, USNO, and VLBA2cm1 images.  

\smallskip\noindent {\bf J0920$+$4441:} This is a complex source.
We have assumed the core to be the strong, 
compact, central component. There is extended
emission on both sides of this core.
The area subtended by the core is $<0.3~\rm{mas}^2 $.

\smallskip\noindent {\bf J0927$+$3902:} (4C39.25)
An image made from VSOP GOT data can be found
in VSOPPR. The main component is slightly
asymmetric, with a small component that is
east of the main body of the source. Higher frequency USNO images
suggest that the core is the more central, stronger component.



\smallskip\noindent {\bf J1058$+$0133:} See
VLBApls, USNO ,VLBA2cm1, and VLBA2cm2 for more detailed images
of the extended emission to the northwest.
Based upon the USNO higher frequency images, the core appears to be
the easternmost component. 

\smallskip\noindent {\bf J1107$-$4449:} The image from the complete
GOT data-set is given in VSOPsth.

\smallskip\noindent {\bf J1118$+$1234:} See VLBA2cm2 for a more 
detailed image.
The area subtended by the core is $<0.1~\rm{mas}^2 $.

\smallskip\noindent {\bf J1146$-$2447:} 
This is a GPS source \citep{sta97}.
The secondary component to the south coincides with the extended
emission seen in the VLBApls and  8~GHz USNO images.

\smallskip\noindent {\bf J1147$-$0724:} The core region extends
westward towards the extended emission seen in the
VLBApls, USNO and VLBA2cm1 images.

\smallskip\noindent {\bf J1147$-$3812:}
Faint emission to the east is also seen in some of the 8~GHz USNO images.

\smallskip\noindent {\bf J1215$-$1731:}
\cite{sti96} were able to get an optical ID for this source, despite
the close proximity of the star gamma
Corvus.

\smallskip\noindent {\bf J1229$+$0203:} (3C273)
A VSOP image made from GOT data
can be found in \cite{lob01}. The four component
model-fit does not include the large-scale structure.

\smallskip\noindent {\bf J1230$+$1223:} (3C274, M87)
A VSOP image made from GOT data 
can be found in \cite{jun00}.  The four component
model fit only accounts for the small-scale structure. The faint core
is unresolved subtending an area of $<0.05~\rm{mas}^2 $. 

\smallskip\noindent {\bf J1246$-$2547:} The extension from the core
is in the same direction as the extended emission seen in the VLBApls
and VLBA2cm2 images.

\smallskip\noindent {\bf J1256$-$0547:} (3C279) 
For the GOT images of this source see \cite{pin00}.
The core has been identified as the stronger, easternmost component
\citep{pin00}.

\smallskip\noindent {\bf J1310$+$3220:} Based upon 8, 24, and 43~GHz
USNO images we have assumed the core to be the more easterly 
component.

\smallskip\noindent {\bf J1337$-$1257:}
The faint component to the west of the main component is seen
in VLBA2cm1 images after 1999, so it is probably a new
component.

\smallskip\noindent {\bf J1357$-$1744:}
A VLBA image at 5~GHz \citep{fre97} shows a weak, 10-mas
scale
bent jet extending towards the southeast and east.

\smallskip\noindent {\bf J1407$+$2827:} (OQ208)
The area subtended by the faint core is $<0.04~\rm{mas}^2 $.


\smallskip\noindent {\bf J1507$-$1652:}
VLBA2cm1 images, as well as the 8 and 15~GHz USNO images, show
extended structure to the south and southeast. 
This extended structure is not as apparent in our image
due to poor sampling of the short $(u,v)$ spacings.

\smallskip\noindent {\bf J1510$-$0543:}
The extended emission to the east seen in the VLBApls, VLBA2cm1, and
VLBA2cm2 images is not seen in our image, probably due to the
paucity of data on shorter baselines.

\smallskip\noindent {\bf J1512$-$0905:}
The redshift for this source is from \cite{tho90}. 
Optical ID is from \cite{hew93}.


\smallskip\noindent {\bf J1549$+$0237:}
A 5~GHz GOT image in 2000 \citep{mos02} shows a jet
extending to the south. The source underwent an outburst
between 1996 and 1999 \citep{tin03}.

\smallskip\noindent {\bf J1613$+$3412:}
Due to the lack of data on shorter baselines
the extended emission to the
south, seen in the VLBA2cm1 and USNO images, is not seen in our image.
The area subtended by the core is $<0.07~\rm{mas}^2 $.

\smallskip\noindent {\bf J1617$-$7717:}
An image from the complete GOT data-set
is given in VSOPsth.
They identify the core as the weaker, more compact component
to the east.
The area subtended by the faint core is $<0.9~\rm{mas}^2 $.

\smallskip\noindent {\bf J1626$-$2951:}
See VSOPsth for the image with all the GOT data.

\smallskip\noindent {\bf J1635$+$3808:} See VSOPPR for the
image using the entire GOT data-set.
The area subtended by the faint core is $<0.08~\rm{mas}^2 $.

\smallskip\noindent {\bf J1638$+$5720:} 
The core has been identified as the more northerly component in VSOPPR,
where an image using the entire GOT data-set can be found.
The area subtended by the core is $<0.4~\rm{mas}^2 $.


\smallskip\noindent {\bf J1642$+$3948:} (3C345) A GOT image
for this source can be found in VSOPPR.

\smallskip\noindent {\bf J1642$+$6856:} An image with the full GOT
data-set is given in
VSOPPR. In our image, the more northerly emission has been modeled
by two unresolved components.
As in VSOPPR, we have assumed the core is more northerly of these
two components.
The area subtended by this core component is $<0.2~\rm{mas}^2 $.

\smallskip\noindent {\bf J1653$+$3945:} (Mkn 501) The GOT
image for this
source can be found in
\cite{edw00}.

\smallskip\noindent {\bf J1658$+$0515:} This source consists of two
closely spaced components, and is similar to that found in the VLBA2cm1. 
Since the extended
emission is to the east (VLBApls and USNO), the core is probably the
component to the west. 

\smallskip\noindent {\bf J1658$+$0741:} The emission near the source
center contains two components. Presumably, the component to the
south-east is the core. VLBApls, USNO, VLBA2cm1, and VLBA2cm2 images
show a jet linking the core to the secondary component.

\smallskip\noindent {\bf J1723$-$6500:} This is an extremely
low redshift GPS galaxy \citep{tin97}. Since the core of GPS
galaxies are usually weak, we have assumed that the core for
this source was undetected during the observation.
Both components in our image
are of similar strength.

\smallskip\noindent {\bf J1733$-$1304:} (NRAO530)
There are no data on the short spacings, so the
large-scale structure seen in other VLBI images is missing.

\smallskip\noindent {\bf J1740$+$5221:} 
A GOT image can be found in VSOPPR.


\smallskip\noindent {\bf J1800$+$7828:} A GOT image can be
found in VSOPPR.

\smallskip\noindent {\bf J1806$+$6949:} (3C371) A GOT image can be
found in VSOPPR. 

\smallskip\noindent {\bf J1824$+$5651:} 
A GOT image can be found in VSOPPR.

\smallskip\noindent {\bf J1902$+$3159:} (3C395)
The core has been identified as the most westerly component by \cite{lar97}.

\smallskip\noindent {\bf J1924$-$2914:} 
The $(u,v)$ plane coverage for this source is poor, so the much of the 
extended emission to the northeast is missing.
For more detailed images of the extended emission
see \cite{she99}, VLBApls, USNO, VLBA2cm1, and VLBA2cm2.
The area subtended by the faint core is $<0.1~\rm{mas}^2 $.

\smallskip\noindent {\bf J1939$-$1525:}
The image found in VLBA2cm2 has an extra component to the east
of the core.


\smallskip\noindent {\bf J2005$+$7752:}
The extended emission
in the VLBApls, USNO, VLBA2cm1, and VLBA2cm2 images extends more directly
towards the east.

\smallskip\noindent {\bf J2022$+$6136:} This is a GPS source.
Images of this source made from GOT data can be found in 
VSOPPR and \cite{tsc00}. The core identification is
from the latter reference.

\smallskip\noindent {\bf J2101$+$0341:}
The VLBA2cm2 image has a jet to the north-northeast,
not seen in our image.
The area subtended the unresolved core is $<0.07~\rm{mas}^2 $.

\smallskip\noindent {\bf J2129$-$1538:} This is a GPS source \citep{sta97}.
The area subtended by the faint core is $<0.05~\rm{mas}^2 $.

\smallskip\noindent {\bf J2136$+$0041:} This is a complex source.
The structure is
similar to that of the USNO and VLBA2cm1 images.
Since most of the extended emission lies to the west we
have assumed the core to be the easternmost component.

\smallskip\noindent {\bf J2158$-$1501:}
The area subtended by the faint core is $<0.2~\rm{mas}^2 $.

\smallskip\noindent {\bf J2202$+$4216:} (BL Lac)
A VSOP image made from GOT data can be found in \cite{oka00}.

\smallskip\noindent {\bf J2212$+$2355:} The
data for this source are of
poor quality, and the image has a
relatively low resolution.
The optical ID is from \cite{con83}.


\smallskip\noindent {\bf J2229$-$0832:} 
Other VLBI images have an
additional 1.2 Jy of integrated flux density, as well as
extended emission to the north.
ATCA monitoring \citep{tin03}
indicates this source is highly variable at 4.8~GHz.

\smallskip\noindent {\bf J2253$+$1608:} (3C454.3) 
Although some of the large-scale emission is absent, 
our image is consistent with the bent-jet morphology seen
in other VLBI images.
Our model for this source includes VLBApls components 1,3 and 4.

\smallskip\noindent {\bf J2320$+$0513:} 
Our image does not show the extended emission to the northwest seen
in other VLBI images.
The core region contains two
components.
The component to the
south-east is probably the core.

\smallskip\noindent {\bf J2329$-$4730:}
The area subtended by this unresolved component is $<0.1~\rm{mas}^2$.

\smallskip\noindent {\bf J2331$-$1556:}
The area subtended by the core is $<0.2~\rm{mas}^2$.

\smallskip\noindent {\bf J2348$-$1631:} The data quality for this source
is poor, but confirms the overall size and shape of the source found in
the VLBApls, USNO and VLBA2cm1 images. 
The area subtended by the core is $<0.3~\rm{mas}^2 $.

\normalsize

\section{DISCUSSION}

Of the 98 sources for which we were able to determine a
source frame core brightness temperature,
53 had brightness temperatures 
in excess of the canonical inverse Compton limit of
$10^{12}$~K. This is the theoretical upper limit on the brightness
temperature for sources radiating 
incoherent synchrotron radiation \citep{kel69}.
Above this limit inverse Compton scattering leads to rapid cooling of the
electron-photon plasma. Hence,
explaining the high brightness
temperatures
requires other mechanisms, such as relativistic Doppler beaming.
The maximum source frame brightness temperature we observed
was $1.2\times10^{13}$~K
(for the source J0539$-$2839).
This is below the highest core brightness temperature detected to date of
$5.8 \times 10^{13}$~K (for the BL Lac object AO 0235+164, \cite{fre00}),
but is well above the inverse Compton limit.
The median source frame core brightness temperature of
$\approx 10^{12}$~K is in good
agreement with the preliminary results found by \cite{lov00a}.  

A statistical analysis of a somewhat larger sample of the VSOP
Survey data is given by \cite{hor04}, who also model the angular
size and brightness temperature distributions.  This larger sample includes 
those sources which were found to be too resolved to be
included in the VSOP source list.  The brightness temperature and
angular size distribution found at 15 GHz (Y. Y. Kovalev et al., in preparation)
is also
similar to those reported in this paper.

    Because of the variability of many of the sources in the Survey
sample, detailed spectral indices of the core components are difficult
to determine. However, many of the sources were observed with the
VLBA at 15 GHz as part of the VLBA2cm2 survey,
and the
spectral properties of the cores will be reported elsewhere
(L. I. Gurvits et al.,
in preparation). 

    In conclusion, we find that about half of the AGN sample of sources
reported upon in this paper have significant radio 
emission in the core component,
with $T_b\approx 10^{12}$~K at 5~GHz in the source frame.
Since the maximum brightness temperature one is able to determine
using only ground-based arrays is of the order of $10^{12}$~K,
our results confirm the necessity of using space VLBI
to explore the extremely high brightness temperature regime.
In addition, 
our Survey results clearly show that by using
space VLBI with higher sensitivity,
and somewhat higher resolution,
the radio cores of many
AGN can be successfully imaged.

\acknowledgements

We gratefully acknowledge the VSOP Project, which is led by the Institute of Space and Astronautical Science 
of the Japan Aerospace Exploration Agency,
in
cooperation with many organizations and radio telescopes around the world. 
WKS and ART wish to acknowledge support from the Canadian Space Agency.
JEJL and GAM acknowledge support from the Japan Society for the
Promotion of Science.
JEJL is also thankful for support from the
Australian Commonwealth Scientific \& Industrial Research Organization.
RD is supported by a grant from the Japan Society for the
Promotion of Science.
SH acknowledges support through an NRC/NASA-JPL Research
Associateship. SF acknowledges the Bolyai Scholarship received from the
Hungarian Academy of Sciences.
This research has made use of data from the University of Michigan 
Radio Astronomy Observatory which is 
supported by funds from the University of Michigan,
the United States Naval Observatory
(USNO) Radio Reference Frame Image Database (RRFID), and
the NASA/IPAC Extragalactic Database (NED) which is 
operated by the Jet Propulsion Laboratory, California Institute
of Technology, under contract with the National 
Aeronautics and Space Administration. 
The NRAO is a facility of the National Science Foundation, operated
under cooperative agreement by Associated Universities, Inc.
The Australia Telescope
Compact Array is part of the Australia Telescope which is funded by the
Commonwealth of Australia for operation as a National Facility managed
by CSIRO.

\clearpage
\begin{landscape}
\begin{deluxetable}{cllcllccclll}
\tabletypesize{\scriptsize}
\tablecaption{Survey Experiment Details\label{tab1}}
\tablewidth{0pt}
\tablehead{
\multicolumn{2}{c}{Source Names} &
\colhead{Obs.} &
\colhead{Obs. Date} &
\colhead{GRTs} &
\colhead{TSs} &
\colhead{Time On} &
\colhead{HALCA Time} &
\colhead{Corr.} &
\colhead{ID} &
\colhead{$z$} &
\colhead{Refers.} \\
\colhead{J2000} &
\colhead{B1950} &
\colhead{Code} &
\colhead{} &
\colhead{} &
\colhead{} &
\colhead{Src.} &
\colhead{On Src.} &
\colhead{} &
\colhead{} &
\colhead{} &
\colhead{} \\
\colhead{} &
\colhead{} &
\colhead{} &
\colhead{(dd-mm-yyyy)} &
\colhead{} &
\colhead{} &
\colhead{(hh:mm)} &
\colhead{(hh:mm)} &
\colhead{} &
\colhead{} &
\colhead{} &
\colhead{} 
}
\startdata
 J0006$-$0623 & 0003$-$066 & vs03t &29-08-1998&THM    &R   &  2:15 &  2:00 & P& B&0.347&123    \\ 
 J0019$+$7327 & 0016$+$731 & vs07a &02-03-1998&lms    &GN  &  4:30 &  2:30 & S& Q&1.781&12357    \\ 
 J0042$+$2320 & 0039$+$230 & vs07u &12-08-1999&MS     &N   &  2:30 &  2:30 & P& E&\phm{1.}$\cdots$\phm{0}&124   \\ 
 J0106$-$4034 & 0104$-$408 & vs03s &08-06-1998&HM     &R   &  3:30 &  3:30 & P& Q&0.584&12    \\ 
 J0115$-$0127 & 0112$-$017 & vs07t &10-08-1999&TMS    &GRT &  3:45 &  3:45 & P& Q&1.365&1234   \\ 
 J0121$+$1149 & 0119$+$115 & vs08p &22-07-2001&RS     &R   &  2:30 &  2:30 & P& Q&0.570&123    \\ 
 J0126$+$2559 & 0123$+$257 & vs08o &20-07-2001&RS     &U   &  3:15 &  3:15 & P& Q&2.370&12    \\ 
 J0136$+$4751 & 0133$+$476 & vs03r &16-08-1999&mnps   &R   &  4:30 &  4:30 & S& Q&0.859&123457   \\ 
 J0210$-$5101 & 0208$-$512 & vs03a &10-06-1998&ATm    &TG  &  4:15 &  4:15 & P& B&1.003&6    \\ 
 J0217$+$7349 & 0212$+$735 & vs02o &05-09-1997&pmh    &N   &  2:15 &  2:15 & S& Q&2.367&12357    \\ 
 J0251$+$4315 & 0248$+$430 & vs07q &15-02-1999&EJP    &NR  &  0:45 &  0:45 & S& Q&1.310&124   \\ 
 J0319$+$4130 & 3C84       & vs01c &25-08-1998&hkms   &GT  &  4:00 &  4:00 & S& G&0.017&157    \\ 
 J0334$-$4008 & 0332$-$403 & vs04b &10-07-1998&HU     &R   &  3:00 &  3:00 & P& B&1.445*&1    \\ 
 J0348$-$2749 & 0346$-$279 & vs08m &09-08-2001&CMS    &U   &  2:00 &  1:30 & P& Q&0.987&1    \\ 
 J0403$-$3605 & 0402$-$362 & vs03z &30-07-1998&CH     &N   &  1:30 &  1:30 & P& Q&1.417&4   \\ 
 J0405$-$1308 & 0403$-$132 & vs03e &19-08-1998&MS     &GR  &  3:45 &  3:15 & P& Q&0.571&124   \\ 
 J0423$-$0120 & 0420$-$014 & vs02g &04-02-1999&bhp    &T   &  4:00 &  4:00 & S& Q&0.915&123    \\ 
 J0440$-$4333 & 0438$-$436 & vs01r &07-03-1998&AHM    &R   &  3:30 &  3:30 & M& Q&2.852&146   \\ 
 J0453$-$2807 & 0451$-$282 & vs04g &12-09-1999&KM     &G   &  3:30 &  3:30 & M& Q&2.560&1    \\ 
 J0457$-$2324 & 0454$-$234 & vs05n &27-02-2002&KMT    &U   &  3:45 &  1:00 & P& Q&1.003&14   \\ 
 J0501$-$0159 & 0458$-$020 & vs02z &22-09-1999&AKM    &R   &  2:15 &  2:15 & P& Q&2.286&123    \\ 
 J0538$-$4405 & 0537$-$441 & vs02c &01-03-1998&HTM    &N   &  2:30 &  2:30 & M& Q&0.896&26    \\ 
 J0539$-$2839 & 0537$-$286 & vs10g &01-10-1999&AHTM   &GU  &  3:00 &  2:15 & P& Q&3.104&14   \\ 
 J0542$+$4951 & 3C147      & vs01p &18-03-1999&RSPU   &NTU &  8:00 &  5:30 & M& Q&0.545&12    \\ 
 J0555$+$3948 & 0552$+$398 & vs01n &23-03-1999&bnS    &G   &  2:15 &  2:15 & S& Q&2.363&1234   \\ 
 J0607$-$0834 & 0605$-$085 & vs03p &14-01-1999&HTKMS  &RT  &  7:45 &  5:00 & P& Q&0.872&123    \\ 
 J0609$-$1542 & 0607$-$157 & vs02a &04-04-1998&HTMSU  &R   &  4:30 &  0:30 & P& Q&0.324&123    \\ 
 J0635$-$7516 & 0637$-$752 & vs01t &21-11-1997&AHM    &G   &  5:00 &  3:45 & P& Q&0.651&6    \\ 
 J0714$+$3534 & 0711$+$356 & vs09k &09-04-1999&hmo    &NT  &  5:00 &  4:00 & S& Q&1.620&12457   \\ 
 J0738$+$1742 & 0735$+$178 & vs05d &30-01-1999&bmps   &G   &  2:45 &  2:30 & S& B&0.424&123    \\ 
 J0741$+$3112 & 0738$+$313 & vs02k &10-01-1999&mns    &T   &  2:30 &  2:30 & S&Q*&0.631*&123    \\ 
 J0748$+$2400 & 0745$+$241 & vs10d &09-02-1999&KMS    &U   &  3:00 &  3:00 & M& G&0.409&123    \\ 
 J0811$+$0146 & 0808$+$019 & vs07n &07-01-1999&bmp    &T   &  2:00 &  2:00 & S& B&0.930*&123    \\ 
 J0818$+$4222 & 0814$+$425 & vs06g &24-04-1999&lms    &NT  &  3:00 &  3:00 & S& B&0.245*&12357    \\ 
 J0824$+$5552 & 0820$+$560 & vs09i &15-10-2000&RKN    &TU  &  2:30 &  2:30 & M& Q&1.417&12    \\ 
 J0836$-$2016 & 0834$-$201 & vs02y &23-01-1999&HTKMS  &RT  &  5:15 &  0:45 & P& Q&2.752&13    \\ 
 J0841$+$7053 & 0836$+$710 & vs04e &07-10-1997&lms    &RT  &  5:30 &  3:45 & S& Q&2.218&12357    \\ 
 J0854$+$2006 & OJ287      & vs03y &04-04-1999&bEfs   &G   &  4:00 &  4:00 & S& B&0.306&123    \\ 
 J0903$+$4651 & 0859$+$470 & vs09g &14-02-1999&EGO    &T   &  5:15 &  1:00 & S& Q&1.462&12357    \\ 
 J0909$+$0121 & 0906$+$015 & vs11r &09-01-1999&TMS    &T   &  4:45 &  3:00 & P& Q&1.018&123    \\ 
 J0920$+$4441 & 0917$+$449 & vs07x &07-02-1999&bfms   &T   &  4:00 &  3:15 & S& Q&2.180&123    \\ 
 J0927$+$3902 & 4C39.25    & vs01f &23-10-1997&msY    &R   &  2:45 &  1:45 & S& Q&0.698&123457   \\ 
 J1037$-$2934 & 1034$-$293 & vs05t &02-06-1999&hos    &G   &  2:45 &  2:45 & S& Q&0.312&124   \\ 
 J1048$+$7143 & 1044$+$719 & vs04j &28-04-1999&SP     &GTU &  4:15 &  3:30 & M& Q&1.150&12    \\ 
 J1058$+$0133 & 1055$+$018 & vs02l &12-05-1999&Ehns   &N   &  3:00 &  3:00 & S& Q&0.888&1234   \\ 
 J1107$-$4449 & 1104$-$445 & vs02v &27-05-1999&ATS    &RU  &  3:15 &  2:00 & P& Q&1.598&6    \\ 
 J1118$+$1234 & 1116$+$128 & vs05s &17-12-1997&KMNSU  &NT  &  4:15 &  2:00 & P& Q&2.118&124   \\ 
 J1146$-$2447 & 1143$-$245 & vs06w &26-05-1999&HTS    &RU  &  7:45 &  6:45 & M& Q&1.940&12    \\ 
 J1147$-$0724 & 1145$-$071 & vs09x &08-03-2000&TKM    &GT  &  5:30 &  4:15 & P& Q&1.342&123    \\ 
 J1147$-$3812 & 1144$-$379 & vs04d &28-12-1997&TmM    &G   &  3:45 &  3:45 & P& Q&1.048&12    \\ 
 J1215$-$1731 & 1213$-$172 & vs05a &11-01-1998&GHN    &RT  &  3:30 &  0:45 & M&G*&\phm{1.}$\cdots$\phm{0}&124   \\ 
 J1229$+$0203 & 3C273B     & vs01b &22-12-1997&blmns  &NR  &  4:15 &  4:15 & S& Q&0.158&13    \\ 
 J1230$+$1223 & 3C274,M87  & vs01a &20-12-1997&bmps   &NR  &  4:00 &  3:30 & S& G&0.004&123    \\ 
 J1246$-$2547 & 1244$-$255 & vs04r &21-01-1998&HTMSU  &GT  &  2:45 &  2:45 & P& Q&0.638&14   \\ 
 J1256$-$0547 & 3C279      & vs01g &10-01-1998&kms    &R   &  2:00 &  2:00 & S& Q&0.538&123    \\ 
 J1310$+$3220 & 1308$+$326 & vs02e &29-06-1998&hns    &T   &  1:30 &  1:00 & S& Q&0.997&1234   \\ 
 J1337$-$1257 & 1334$-$127 & vs01y &10-07-1999&bfms   &T   &  3:15 &  2:30 & S& Q&0.539&123    \\ 
 J1357$-$1744 & 1354$-$174 & vs08j &30-01-1998&AHMU   &GT  &  4:45 &  2:30 & P& Q&3.147&14   \\ 
 J1407$+$2827 & OQ208      & vs03o &30-06-1998&Ehs    &T   &  1:15 &  1:15 & S& G&0.077&123    \\ 
 J1430$+$1043 & 1427$+$109 & vs09c &18-01-2001&CHMNS  &T   &  7:15 &  3:15 & P& Q&1.710&1    \\ 
 J1507$-$1652 & 1504$-$166 & vs03m &10-04-1998&HMU    &NT  &  4:30 &  1:45 & P& Q&0.876&123    \\ 
 J1510$-$0543 & 1508$-$055 & vs06t &16-04-1998&TU     &GNT &  1:45 &  1:45 & P& Q&1.191&134   \\ 
 J1512$-$0905 & 1510$-$089 & vs02x &11-08-1999&mos    &GT  &  2:30 &  2:30 & S&Q*&0.360*&1234   \\ 
 J1517$-$2422 & 1514$-$241 & vs03u &27-04-1998&HTU    &GNU &  7:30 &  4:30 & P& B&0.048&123    \\ 
 J1549$+$0237 & 1546$+$027 & vs06e &31-07-1998&GH     &RU  &  3:00 &  2:00 & P& Q&0.412&1234   \\ 
 J1613$+$3412 & 1611$+$343 & vs02b &04-02-1998&mn     &GT  &  7:00 &  6:00 & M& Q&1.401&123    \\ 
 J1617$-$7717 & 1610$-$771 & vs02u &05-04-1999&AM     &T   &  3:30 &  3:00 & P& Q&1.710&6    \\ 
 J1626$-$2951 & 1622$-$297 & vs03l &22-02-1998&HTS    &T   &  1:15 &  1:15 & P& Q&0.815&146   \\ 
 J1635$+$3808 & 1633$+$382 & vs03d &04-08-1998&bmns   &RT  &  4:30 &  4:30 & S& Q&1.807&12357    \\ 
 J1638$+$5720 & 1637$+$574 & vs06n &21-04-1998&hlm    &N   &  4:00 &  4:00 & S& Q&0.751&1257    \\ 
 J1640$+$3946 & NRAO512    & vs08y &03-09-1999&KN     &R   &  1:00 &  1:00 & P& Q&1.666&1234   \\ 
 J1642$+$3948 & 3C345      & vs01k &28-07-1998&bmns   &RT  &  3:45 &  3:15 & S& Q&0.594&1235    \\ 
 J1642$+$6856 & 1642$+$690 & vs07w &31-05-1998&ENS    &GT  &  5:45 &  4:15 & S& G&0.751&12357    \\ 
 J1653$+$3945 & Mkn 501    & vs08h &07-04-1998&hlm    &NR  &  3:15 &  3:15 & S& B&0.033&12357    \\ 
 J1658$+$0515 & 1656$+$053 & vs05i &04-03-1998&HTMS   &GT  &  4:00 &  8:45 & P& Q&0.879&1234   \\ 
 J1658$+$0741 & 1655$+$077 & vs07g &05-03-1998&HTMNS  &GNT &  7:15 &  2:30 & P& Q&0.621&1234   \\ 
 J1723$-$6500 & 1718$-$649 & vs02f &25-03-1999&AHT    &NT  &  3:00 &  0:15 & P& G&0.014&6    \\ 
 J1733$-$1304 & NRAO530    & vs01m &08-09-1997&AMU    &U   &  3:30 &  3:30 & P& Q&0.902&123    \\ 
 J1740$+$5211 & 1739$+$522 & vs10v &14-06-1998&mns    &NR  &  3:30 &  3:30 & S& Q&1.379&12357    \\ 
 J1751$+$0939 & 1749$+$096 & vs04o &20-08-1998&hmn    &R   &  2:30 &  2:30 & S& Q&0.320&123    \\ 
 J1800$+$7828 & 1803$+$784 & vs02w &16-10-1997&hlm    &R   &  3:00 &  3:00 & S& Q&0.680&12357    \\ 
 J1806$+$6949 & 3C371      & vs04x &11-03-1998&hmns   &NT  &  4:30 &  4:30 & S& B&0.050&12357    \\ 
 J1824$+$5651 & 1823$+$568 & vs06d &31-05-1998&EGN    &R   &  4:30 &  2:30 & S& Q&0.663&12357    \\ 
 J1902$+$3159 & 3C395      & vs06c &01-05-1998&hnps   &N   &  4:30 &  4:30 & S& Q&0.635&123    \\ 
 J1924$-$2914 & 1921$-$293 & vs01e &19-06-1998&los    &R   &  1:45 &  1:45 & S& Q&0.352&1234   \\ 
 J1939$-$1525 & 1936$-$155 & vs06m &22-07-1998&HTS    &GT  &  4:30 &  3:15 & P& Q&1.657&124   \\ 
 J2000$-$1748 & 1958$-$179 & vs04w &25-06-1998&TS     &NT  &  2:30 &  2:30 & P& Q&0.652&124   \\ 
 J2005$+$7752 & 2007$+$777 & vs06r &10-03-1998&bfms   &G   &  3:00 &  2:00 & S& B&0.342&1234   \\ 
 J2022$+$6136 & 2021$+$614 & vs02q &06-11-1997&hkm    &T   &  4:15 &  4:15 & S& G&0.227&123457   \\ 
 J2101$+$0341 & 2059$+$034 & vs08v &14-11-2000&KS     &R   &  1:45 &  1:45 & P& Q&1.015&124   \\ 
 J2129$-$1538 & 2126$-$158 & vs08e &13-07-1998&fmnps  &GR  &  3:00 &  3:00 & S& Q&3.280&124   \\ 
 J2136$+$0041 & 2134$+$004 & vs01h &28-11-1997&TMU    &T   &  3:45 &  2:15 & P& Q&1.932&123    \\ 
 J2158$-$1501 & 2155$-$152 & vs03g &15-08-1998&HTMS   &T   &  3:00 &  1:00 & P& Q&0.672&123    \\ 
 J2202$+$4216 & BL Lac     & vs01q &08-12-1997&lms    &T   &  5:15 &  4:30 & S& B&0.069&123457   \\ 
 J2212$+$2355 & 2209$+$236 & vs06l &27-05-1998&XGH    &R   &  1:30 &  1:30 & P&Q*&\phm{1.}$\cdots$\phm{0}&    \\ 
 J2225$-$0457 & 3C446      & vs01s &02-12-2000&CHS    &T   &  5:30 &  3:45 & P& Q&1.404&123    \\ 
 J2229$-$0832 & 2227$-$088 & vs04i &27-11-1997&HTKMS  &RT  &  6:45 &  3:15 & P& Q&1.562&1234   \\ 
 J2253$+$1608 & 3C454.3    & vs01d &12-12-1997&nps    &N   &  1:15 &  1:15 & S& Q&0.859&123    \\ 
 J2320$+$0513 & 2318$+$049 & vs09p &08-12-2000&CHMNS  &NRT &  7:30 &  5:30 & P& Q&0.623&1234   \\ 
 J2329$-$4730 & 2326$-$477 & vs04t &19-06-1998&HM     &R   &  4:30 &  3:30 & P& Q&1.306&    \\ 
 J2331$-$1556 & 2329$-$162 & vs06b &11-06-1998&HTK    &GU  &  3:30 &  2:00 & P& Q&1.153&124   \\ 
 J2348$-$1631 & 2345$-$167 & vs03b &08-06-1998&TU     &G   &  4:00 &  3:30 & P& Q&0.576&123    \\ 
\enddata
\tablecomments{
Col. $(4)$: Starting date of the observation\newline
Col. $(5)$: Ground radio telescopes
to which fringes were successfully obtained. Lower case letters indicate
VLBA GRTs: b$=$Brewster, f$=$Fort Davis, h$=$Hancock, k$=$Kitt Peak, l$=$Los Alamos, m$=$Mauna Kea,
n$=$North Liberty, o$=$Owens Valley, p$=$Pie Town,
s$=$St. Croix. Upper Case letters indicate other GRTs: X$=$Arecibo, A$=$ATCA, C$=$Ceduna, E$=$Effelsburg,
G$=$Green Bank $140^\prime$, H$=$Hartebeesthoek, T$=$Hobart, J$=$Jodrell Bank MKII, R$=$Kalyazin, K$=$Kashima, N$=$Noto,
M$=$Mopra, O$=$Onsala, S$=$Sheshan, P$=$Torun, U$=$Usuda, Y$=$VLA\newline
Col. $(6)$: Tracking Stations to which fringes were successfully obtained:
G$=$Goldstone, N$=$Green Bank, R$=$Robledo,
T$=$Tidbinbilla, U$=$Usuda\newline
Col. $(7)$: Approximate time on source over which fringes were found\newline
Col. $(8)$: Approximate time on source over which fringes were found on space baselines\newline
Col. $(9)$: Correlator: M$=$Mitaka, P$=$Penticton, S$=$Socorro\newline
Col. $(10)$:Optical classification:
Q~--~quasar, B~--~ BL Lac object,
G~--~AGN other 
than B and Q (e.g. Seyfert galaxy),
E~--~empty field or unidentified optical counterpart;
from \cite{ver01} unless appended by an asterisk (*), whereby the
reference is given in the individual source notes\newline
Col. $(11)$: Redshift; from \cite{ver01} unless appended by an
asterisk (*), whereby the reference is given in the source notes.
Col. $(12)$: References: \newline
1 = VLBApls; \citep{fom00b}, http://www.aoc.nrao.edu/vlba/html/6CM/index.htm \newline
2 = USNO; \citep{fey96, fey97, fey00}, http://rorf.usno.navy.mil/RRFID \newline
3 = VLBA2cm1; \citep{kel98, zen02}, http://www.cv.nrao.edu/2cmsurvey/maps/index.html \newline
4 = VLBA2cm2; (L. I. Gurvits, K. I. Kellermann,E. B. Fomalont, 
    \& H. Y. Zhang, in preparation)\newline
5 = VSOPPR; \citep{lis01} \newline
6 = VSOPsouth; \citep{tin02} \newline
7 = PR; \citep{pea88} http://astro.caltech.edu/$\sim$tjp/cj
}
\end{deluxetable}
\end{landscape}
\begin{landscape}
\begin {deluxetable}{lrrrrrrclrrrrrrc}
\tablecaption{Source Component Parameters}
\tabletypesize{\scriptsize}
\tablewidth{0cm}
\tablehead{

\colhead{Source}& 
\colhead{S\phm{*}}& 
\colhead{Rad}& 
\colhead{Phi}& 
\colhead{Bmaj}& 
\colhead{Bmin}& 
\colhead{P.A.}& 
\colhead{$T_b \times 10^{12}$}& 

\colhead{Source}& 
\colhead{S\phm{*}}& 
\colhead{Rad}& 
\colhead{Phi}& 
\colhead{Bmaj}& 
\colhead{Bmin}& 
\colhead{P.A.}& 
\colhead{$T_b \times 10^{12}$}\\

\colhead{}&
\colhead{(Jy)\phm{*}}&
\colhead{(mas)}&
\colhead{($^\circ$)}&
\colhead{(mas)}&
\colhead{(mas)}&
\colhead{($^\circ$)}&
\colhead{(K)}&

\colhead{}&
\colhead{(Jy)\phm{*}}&
\colhead{(mas)}&
\colhead{($^\circ$)}&
\colhead{(mas)}&
\colhead{(mas)}&
\colhead{($^\circ$)}&
\colhead{(K)} 
}
\startdata 
J0006$-$0623 &  1.29*\phm{*}&   0.0 & $\cdots$ &  0.5 &  0.0 &  17 & $>$ 0.48 &            &  0.40\phm{*}\phm{*}&   0.3  &   12 &  0.8 &  0.3 & $-$31 & \phm{$>$}       \\
           &  1.50*\phm{*}&   0.6  &  $-$67 &  4.4 &  0.5 & $-$79 & \phm{$>$}       & J0457$-$2324 &  4.21\phm{*}\phm{*}&   0.0 & $\cdots$ &  0.4 &  0.4 & $\cdots$ & \phm{$>$} 1.40  \\
J0019$+$7327 &  0.06*\phm{*}&   0.2  &   94 &  0.0 &  0.0 & $\cdots$ & $>$ 0.17 & J0501$-$0159 &  0.66\phm{*}\phm{*}&   0.1  &  $-$12 &  1.0 &  0.2 & $-$16 & \phm{$>$} 0.16  \\
           &  0.47*\phm{*}&   0.1  &  152 &  1.1 &  0.6 & $-$61 & \phm{$>$}       &            &  0.04\phm{*}\phm{*}&   1.0  &  $-$28 &  0.0 &  0.0 & $\cdots$ & \phm{$>$}       \\
J0042$+$2320 &  0.29\phm{*}\phm{*}&   0.0 & $\cdots$ &  0.2 &  0.2 & $\cdots$ & \phm{$>$} 0.29  & J0538$-$4405 &  0.57\phm{*}\phm{*}&   0.0 & $\cdots$ &  0.2 &  0.2 & $\cdots$ & \phm{$>$} 0.82  \\
J0106$-$4034 &  2.93\phm{*}\phm{*}&   0.0 & $\cdots$ &  0.3 &  0.0 &  $-$5 & $>$ 3.79 &            &  0.42\phm{*}\phm{*}&   0.8  &   72 &  0.5 &  0.5 & $\cdots$ & \phm{$>$}       \\
J0115$-$0127 &  0.36\phm{*}\phm{*}&   0.1  &  $-$38 &  0.6 &  0.1 & $-$37 & \phm{$>$} 0.25  & J0539$-$2839 &  1.63\phm{*}\phm{*}&   0.0 & $\cdots$ &  0.2 &  0.1 &  84 & \phm{$>$} 3.47  \\
           &  0.55\phm{*}\phm{*}&   1.0  &  114 &  1.4 &  0.1 & $-$53 & \phm{$>$}       & J0542$+$4951 &  0.14\phm{*}\phm{*}&   0.0 & $\cdots$ &  0.0 &  0.0 & $\cdots$ & $>$ 0.07 \\
J0121$+$1149 &  1.92\phm{*}\phm{*}&   0.0 & $\cdots$ &  1.5 &  0.3 &  $-$1 & \phm{$>$} 0.27  &            &  0.36\phm{*}\phm{*}&   1.0  & $-$126 &  1.7 &  1.3 &   3 & \phm{$>$}       \\
J0126$+$2559 &  0.60\phm{*}\phm{*}&   0.2  &  165 &  0.0 &  0.0 & $\cdots$ & $>$ 0.04 & J0555$+$3948 &  4.72\phm{*}\phm{*}&   0.0 & $\cdots$ &  0.4 &  0.4 & $\cdots$ & \phm{$>$} 1.63  \\
           &  0.16\phm{*}\phm{*}&   1.9  & $-$175 &  0.0 &  0.0 & $\cdots$ & \phm{$>$}       &            &  2.57\phm{*}\phm{*}&   0.7  &  $-$53 &  0.6 &  0.6 & $\cdots$ & \phm{$>$}       \\
J0136$+$4751 &  1.82\phm{*}\phm{*}&   0.0 & $\cdots$ &  0.2 &  0.1 & $-$14 & \phm{$>$} 4.27  & J0607$-$0834 &  0.78\phm{*}\phm{*}&   0.0 & $\cdots$ &  0.3 &  0.1 & $-$51 & \phm{$>$} 1.25  \\
           &  0.36\phm{*}\phm{*}&   0.6  &  $-$33 &  1.2 &  0.3 &  $-$3 & \phm{$>$}       &            &  0.41\phm{*}\phm{*}&   1.1  &  126 &  0.6 &  0.6 & $\cdots$ & \phm{$>$}       \\
J0210$-$5101 &  0.72\phm{*}\phm{*}&   0.1  & $-$100 &  0.4 &  0.2 &  51 & \phm{$>$} 0.46  &            &  0.16\phm{*}\phm{*}&   2.0  &  103 &  0.9 &  0.9 & $\cdots$ & \phm{$>$}       \\
           &  1.83\phm{*}\phm{*}&   1.9  & $-$111 &  0.8 &  0.7 & $-$24 & \phm{$>$}       & J0609$-$1542 &  2.13\phm{*}\phm{*}&   0.0 & $\cdots$ &  0.0 &  0.0 & $\cdots$ & $>$ 3.79 \\
J0217$+$7349 &  0.47*\phm{*}&   0.2  &   50 &  0.2 &  0.2 & $\cdots$ & \phm{$>$} 0.50  &            &  0.65\phm{*}\phm{*}&   0.8  &   58 &  1.0 &  0.2 & $-$18 & \phm{$>$}       \\
           &  2.42*\phm{*}&   0.0 & $\cdots$ &  0.8 &  0.3 & $-$63 & \phm{$>$}       & J0635$-$7516 &  3.50*\phm{*}&   0.1  &  $-$10 &  0.7 &  0.4 & $-$76 & \phm{$>$} 0.71  \\
J0251$+$4315 &  0.57\phm{*}\phm{*}&   0.0 & $\cdots$ &  0.4 &  0.4 & $\cdots$ & \phm{$>$} 0.20  &            &  5.68\phm{*}\phm{*}&   1.7  &  $-$87 &  1.1 &  1.1 & $\cdots$ & \phm{$>$}       \\
           &  0.27\phm{*}\phm{*}&   1.7  &  143 &  0.3 &  0.3 & $\cdots$ & \phm{$>$}       &            &  1.69*\phm{*}&   0.3  & $-$106 &  0.9 &  0.1 & $-$80 & \phm{$>$}       \\
           &  0.07\phm{*}\phm{*}&   0.9  &  156 &  0.0 &  0.0 & $\cdots$ & \phm{$>$}       & J0714$+$3534 &  0.76\phm{*}\phm{*}&   0.0 & $\cdots$ &  0.7 &  0.5 & $-$65 & \phm{$>$} 0.10  \\
J0319$+$4130 &  1.01\phm{*}\phm{*}&   1.7  &   12 &  0.8 &  0.8 & $\cdots$ & \phm{$>$} 0.08  &            &  0.19\phm{*}\phm{*}&   5.8  &  157 &  0.7 &  0.2 & $-$46 & \phm{$>$}       \\
           &  1.59\phm{*}\phm{*}&   0.3  &   $-$1 &  1.2 &  0.8 & $-$19 & \phm{$>$}       &            &  0.13\phm{*}\phm{*}&   1.5  &  160 &  0.8 &  0.8 & $\cdots$ & \phm{$>$}       \\
           &  1.00\phm{*}\phm{*}&   1.0  &  $-$33 &  2.5 &  0.5 &   3 & \phm{$>$}       & J0738$+$1742 &  0.58\phm{*}\phm{*}&   0.0 & $\cdots$ &  0.2 &  0.2 & $\cdots$ & \phm{$>$} 0.77  \\
           &  0.94\phm{*}\phm{*}&  10.5  &   10 &  1.7 &  1.7 & $\cdots$ & \phm{$>$}       &            &  0.11\phm{*}\phm{*}&   1.0  &   70 &  0.7 &  0.7 & $\cdots$ & \phm{$>$}       \\
           &  0.68\phm{*}\phm{*}&   7.0  & $-$164 &  1.1 &  1.1 & $\cdots$ & \phm{$>$}       & J0741$+$3112 &  3.52\phm{*}\phm{*}&   0.1  &   29 &  0.4 &  0.4 & $\cdots$ & \phm{$>$} 1.05  \\
           &  0.30\phm{*}\phm{*}&   2.4  &  177 &  1.0 &  1.0 & $\cdots$ & \phm{$>$}       &            &  1.27\phm{*}\phm{*}&   2.6  &    0 &  0.4 &  0.4 & $\cdots$ & \phm{$>$}       \\
J0334$-$4008 &  0.49\phm{*}\phm{*}&   0.0 & $\cdots$ &  0.0 &  0.0 & $\cdots$ & $>$ 0.60 & J0748$+$2400 &  0.52\phm{*}\phm{*}&   0.0 & $\cdots$ &  0.6 &  0.0 & $-$63 & $>$ 0.19 \\
J0348$-$2749 &  0.84\phm{*}\phm{*}&   0.0 & $\cdots$ &  0.3 &  0.1 & $-$69 & \phm{$>$} 1.13  &            &  0.11\phm{*}\phm{*}&   3.4  &  $-$60 &  0.3 &  0.3 & $\cdots$ & \phm{$>$}       \\
           &  0.31\phm{*}\phm{*}&   1.2  &  136 &  1.2 &  1.2 & $\cdots$ & \phm{$>$}       & J0811$+$0146 &  0.67\phm{*}\phm{*}&   0.1  &  141 &  0.6 &  0.2 & $-$30 & \phm{$>$} 0.40  \\
J0403$-$3605 &  1.16\phm{*}\phm{*}&   0.0 & $\cdots$ &  0.5 &  0.2 &  21 & \phm{$>$} 0.79  &            &  0.13\phm{*}\phm{*}&   0.7  &  163 &  0.0 &  0.0 & $\cdots$ & \phm{$>$}       \\
           &  0.20\phm{*}\phm{*}&   2.6  &   37 &  0.7 &  0.4 &  10 & \phm{$>$}       & J0818$+$4222 &  0.58\phm{*}\phm{*}&   0.0 & $\cdots$ &  0.2 &  0.1 & $-$74 & \phm{$>$} 1.16  \\
J0405$-$1308 &  0.76\phm{*}\phm{*}&   0.1  &   47 &  0.2 &  0.2 & $\cdots$ & \phm{$>$} 0.98  &            &  0.37\phm{*}\phm{*}&   1.1  &   86 &  0.8 &  0.8 & $\cdots$ & \phm{$>$}       \\
J0423$-$0120 &  2.68\phm{*}\phm{*}&   0.0 & $\cdots$ &  0.3 &  0.3 & $\cdots$ & \phm{$>$} 1.90  & J0824$+$5552 &  0.75\phm{*}\phm{*}&   0.0 & $\cdots$ &  0.4 &  0.2 & $-$74 & \phm{$>$} 0.47  \\
J0440$-$4333 &  0.75*\phm{*}&   0.0 & $\cdots$ &  0.2 &  0.2 & $\cdots$ & \phm{$>$} 1.66  & J0836$-$2016 &  1.64\phm{*}\phm{*}&   0.0 & $\cdots$ &  1.0 &  0.8 &  30 & \phm{$>$} 0.11  \\
           &  0.46*\phm{*}&   0.3  &  105 &  0.6 &  0.6 & $\cdots$ & \phm{$>$}       & J0841$+$7053 &  0.27*\phm{*}&   0.1  &   11 &  0.0 &  0.0 & $\cdots$ & $>$ 0.12 \\
J0453$-$2807 &  0.38\phm{*}\phm{*}&   0.1  &  156 &  0.2 &  0.1 & $-$18 & \phm{$>$} 1.08  &            &  0.66\dag\phm{*}&   2.1  & $-$142 &  2.5 &  0.5 &  25 & \phm{$>$}       \\
           &  0.64*\phm{*}&   0.1  & $-$133 &  0.7 &  0.2 &  34 & \phm{$>$}       &            &  0.42*\phm{*}&   0.9  &  $-$80 &  1.5 &  0.0 &  50 & \phm{$>$}       \\
           &  0.11\dag\phm{*}&   2.6  & $-$140 &  0.0 &  0.0 & $\cdots$ & \phm{$>$}       & J1246$-$2547 &  0.68\phm{*}\phm{*}&   0.1  &   37 &  0.2 &  0.2 & $\cdots$ & \phm{$>$} 0.71  \\
J0854$+$2006 &  0.40\phm{*}\phm{*}&   0.2  &   33 &  0.0 &  0.0 & $\cdots$ & $>$ 0.07 &            &  0.11\phm{*}\phm{*}&   0.6  &  142 &  0.4 &  0.4 & $\cdots$ & \phm{$>$}       \\
           &  1.42\phm{*}\phm{*}&   0.1  &  $-$61 &  0.3 &  0.2 &  57 & \phm{$>$}       & J1256$-$0547 &  7.93\phm{*}\phm{*}&   0.2  & $-$145 &  0.7 &  0.3 &  40 & \phm{$>$} 2.02  \\
           &  0.30\phm{*}\phm{*}&   0.9  &  $-$91 &  0.4 &  0.2 &  60 & \phm{$>$}       &            &  5.75\phm{*}\phm{*}&   3.2  & $-$113 &  1.3 &  0.3 &  11 & \phm{$>$}       \\
J0903$+$4651 &  0.39*\phm{*}&   0.1  &  172 &  0.9 &  0.1 & $-$25 & \phm{$>$} 0.25  & J1310$+$3220 &  1.39\phm{*}\phm{*}&   0.0 & $\cdots$ &  0.2 &  0.2 & $\cdots$ & \phm{$>$} 1.25  \\
           &  0.35*\phm{*}&   0.5  &  125 &  1.2 &  1.2 & $\cdots$ & \phm{$>$}       &            &  1.07\phm{*}\phm{*}&   1.1  &  $-$58 &  0.4 &  0.4 & $\cdots$ & \phm{$>$}       \\
           &  0.27\phm{*}\phm{*}&   1.5  &    1 &  2.0 &  1.0 & $-$28 & \phm{$>$}       & J1337$-$1257 &  3.72\phm{*}\phm{*}&   0.0 & $\cdots$ &  0.5 &  0.2 &  13 & \phm{$>$} 1.73  \\
J0909$+$0121 &  0.97\phm{*}\phm{*}&   0.0 & $\cdots$ &  0.4 &  0.2 & $-$23 & \phm{$>$} 0.58  &            &  0.33\phm{*}\phm{*}&   0.7  & $-$128 &  0.0 &  0.0 & $\cdots$ & \phm{$>$}       \\
           &  0.08\phm{*}\phm{*}&   1.0  &   29 &  0.5 &  0.5 & $\cdots$ & \phm{$>$}       & J1357$-$1744 &  0.32*\phm{*}&   0.0 & $\cdots$ &  0.3 &  0.3 & $\cdots$ & \phm{$>$} 0.26  \\
J0920$+$4441 &  0.33\phm{*}\phm{*}&   0.1  &  $-$26 &  0.0 &  0.0 & $\cdots$ & $>$ 0.07 &            &  0.77*\phm{*}&   0.1  & $-$108 &  1.4 &  1.0 &  39 & \phm{$>$}       \\
           &  0.58\phm{*}\phm{*}&   0.9  &  179 &  0.8 &  0.6 &  28 & \phm{$>$}       & J1407$+$2827 &  0.06*\phm{*}&   0.2  &  123 &  0.0 &  0.0 & $\cdots$ & $>$ 0.07 \\
           &  0.28\phm{*}\phm{*}&   0.6  &    7 &  0.5 &  0.3 & $-$70 & \phm{$>$}       &            &  1.89*\phm{*}&   0.0 & $\cdots$ &  1.5 &  0.7 & $-$19 & \phm{$>$}       \\
J0927$+$3902 &  9.67\phm{*}\phm{*}&   0.2  &  136 &  0.5 &  0.5 & $\cdots$ & \phm{$>$} 1.96  & J1430$+$1043 &  0.79\phm{*}\phm{*}&   0.1  & $-$153 &  1.1 &  0.2 &  17 & \phm{$>$} 0.22  \\
           &  0.95\phm{*}\phm{*}&   0.6  &   97 &  0.0 &  0.0 & $\cdots$ & \phm{$>$}       & J1507$-$1652 &  0.93\phm{*}\phm{*}&   0.2  & $-$159 &  0.5 &  0.1 &  26 & \phm{$>$} 0.97  \\
J1037$-$2934 &  1.11\phm{*}\phm{*}&   0.1  & $-$160 &  0.5 &  0.3 & $-$49 & \phm{$>$} 0.33  & J1510$-$0543 &  0.55\phm{*}\phm{*}&   0.0 & $\cdots$ &  0.2 &  0.2 & $\cdots$ & \phm{$>$} 0.53  \\
           &  0.22\phm{*}\phm{*}&   0.7  &  125 &  0.6 &  0.6 & $\cdots$ & \phm{$>$}       & J1512$-$0905 &  1.62\phm{*}\phm{*}&   0.1  &    7 &  0.4 &  0.1 & $-$18 & \phm{$>$} 1.96  \\
J1048$+$7143 &  1.84\phm{*}\phm{*}&   0.0 & $\cdots$ &  0.3 &  0.1 & $-$28 & \phm{$>$} 2.23  &            &  0.28\phm{*}\phm{*}&   1.7  &  $-$39 &  0.9 &  0.9 & $\cdots$ & \phm{$>$}       \\
J1058$+$0133 &  1.02\phm{*}\phm{*}&   0.0 & $\cdots$ &  0.2 &  0.2 & $\cdots$ & \phm{$>$} 2.26  & J1517$-$2422 &  1.04\phm{*}\phm{*}&   0.1  &  170 &  0.9 &  0.1 &  $-$5 & \phm{$>$} 1.11  \\
           &  0.60\phm{*}\phm{*}&   1.8  &  $-$47 &  0.7 &  0.4 & $-$72 & \phm{$>$}       &            &  0.15\phm{*}\phm{*}&   0.9  &  168 &  0.0 &  0.0 & $\cdots$ & \phm{$>$}       \\
           &  0.23\phm{*}\phm{*}&   6.0  &  $-$49 &  2.7 &  1.2 & $-$19 & \phm{$>$}       &            &  0.11\phm{*}\phm{*}&   1.5  &  171 &  0.0 &  0.0 & $\cdots$ & \phm{$>$}       \\
J1107$-$4449 &  2.13\phm{*}\phm{*}&   0.3  &  151 &  0.7 &  0.4 &  74 & \phm{$>$} 0.41  & J1549$+$0237 &  1.59*\phm{*}&   0.0 & $\cdots$ &  0.8 &  0.1 &   0 & \phm{$>$} 1.56  \\
J1118$+$1234 &  0.51\phm{*}\phm{*}&   0.0 & $\cdots$ &  0.6 &  0.0 &  10 & $>$ 0.30 &            &  0.45*\phm{*}&   0.4  &  157 &  1.0 &  0.0 & $-$46 & \phm{$>$}       \\
J1146$-$2447 &  0.44\phm{*}\phm{*}&   0.1  & $-$124 &  0.5 &  0.3 &  32 & \phm{$>$} 0.15  & J1613$+$3412 &  1.05\phm{*}\phm{*}&   0.0 & $\cdots$ &  0.2 &  0.0 &   0 & $>$ 0.95 \\
           &  0.23\phm{*}\phm{*}&   3.8  &  175 &  1.4 &  0.6 & $-$21 & \phm{$>$}       & J1617$-$7717 &  0.24*\phm{*}&   0.2  &  102 &  0.0 &  0.0 & $\cdots$ & $>$ 0.04 \\
J1147$-$0724 &  0.74\phm{*}\phm{*}&   0.0 & $\cdots$ &  0.5 &  0.3 &  21 & \phm{$>$} 0.27  &            &  2.51*\phm{*}&   0.2  &  $-$77 &  1.1 &  0.5 & $-$70 & \phm{$>$}       \\
J1147$-$3812 &  2.08\phm{*}\phm{*}&   0.0 & $\cdots$ &  0.2 &  0.2 & $\cdots$ & \phm{$>$} 1.81  & J1626$-$2951 &  0.59*\phm{*}&   0.1  &   86 &  0.4 &  0.3 &  29 & \phm{$>$} 0.27  \\
J1215$-$1731 &  1.05\phm{*}\phm{*}&   0.0 & $\cdots$ &  0.2 &  0.2 & $\cdots$ & \phm{$>$} 2.11  &            &  1.12*\phm{*}&   0.2  &  $-$75 &  0.7 &  0.4 & $-$72 & \phm{$>$}       \\
           &  0.09\phm{*}\phm{*}&   0.7  &  108 &  0.0 &  0.0 & $\cdots$ & \phm{$>$}       & J1635$+$3808 &  0.42\phm{*}\phm{*}&   0.7  &  107 &  0.0 &  0.0 & $\cdots$ & $>$ 0.30 \\
J1229$+$0203 &  3.07\phm{*}\phm{*}&   0.6  & $-$146 &  0.6 &  0.6 & $\cdots$ & \phm{$>$} 0.50  &            &  0.96\phm{*}\phm{*}&   1.1  &  $-$92 &  1.0 &  0.8 &  64 & \phm{$>$}       \\
           &  5.26\phm{*}\phm{*}&   1.9  & $-$116 &  4.6 &  0.0 &  63 & \phm{$>$}       &            &  0.69\phm{*}\phm{*}&   0.2  &  134 &  0.6 &  0.3 & $-$53 & \phm{$>$}       \\
           &  1.64\phm{*}\phm{*}&   6.3  & $-$110 &  0.6 &  0.6 & $\cdots$ & \phm{$>$}       & J1638$+$5720 &  0.17\phm{*}\phm{*}&   1.0  &   22 &  0.0 &  0.0 & $\cdots$ & $>$ 0.04 \\
           &  0.13\phm{*}\phm{*}&   7.9  & $-$115 &  0.0 &  0.0 & $\cdots$ & \phm{$>$}       &            &  0.29\phm{*}\phm{*}&   0.0 & $\cdots$ &  0.0 &  0.0 & $\cdots$ & \phm{$>$}       \\
J1230$+$1223 &  0.18\phm{*}\phm{*}&   0.7  &   93 &  0.0 &  0.0 & $\cdots$ & $>$ 0.23 & J1640$+$3946 &  1.03\phm{*}\phm{*}&   0.0 & $\cdots$ &  0.4 &  0.3 & $-$47 & \phm{$>$} 0.49  \\
           &  1.61*\phm{*}&   2.5  &  $-$76 &  3.2 &  3.2 & $\cdots$ & \phm{$>$}       & J1642$+$3948 &  1.77\phm{*}\phm{*}&   0.0 & $\cdots$ &  0.4 &  0.2 &  23 & \phm{$>$} 0.99  \\
           &  0.60\phm{*}\phm{*}&   0.4  &   18 &  0.8 &  0.4 & $-$38 & \phm{$>$}       &            &  2.64\phm{*}\phm{*}&   1.4  & $-$100 &  1.4 &  0.0 & $-$81 & \phm{$>$}       \\
           &  2.25\phm{*}\phm{*}&   0.7  & $-$103 &  0.6 &  0.3 & $-$19 & \phm{$>$}       &            &  0.94*\phm{*}&   0.3  & $-$131 &  0.6 &  0.4 & $-$57 & \phm{$>$}       \\
J1642$+$6856 &  0.14\phm{*}\phm{*}&   0.4  &   11 &  0.0 &  0.0 & $\cdots$ & $>$ 0.04 &            &  0.69*\phm{*}&   0.1  &  $-$91 &  0.6 &  0.3 & $-$16 & \phm{$>$}       \\
           &  0.34\phm{*}\phm{*}&   0.2  &   32 &  0.0 &  0.0 & $\cdots$ & \phm{$>$}       &            &  0.14\phm{*}\phm{*}&   9.5  &   41 &  0.9 &  0.3 & $-$36 & \phm{$>$}       \\
           &  0.23\phm{*}\phm{*}&   1.0  &  179 &  0.7 &  0.4 & $-$44 & \phm{$>$}       & J2101$+$0341 &  0.67\phm{*}\phm{*}&   0.0 & $\cdots$ &  0.0 &  0.0 & $\cdots$ & $>$ 0.60 \\
J1653$+$3945 &  0.48\phm{*}\phm{*}&   0.0 & $\cdots$ &  0.2 &  0.2 & $\cdots$ & \phm{$>$} 0.58  & J2129$-$1538 &  0.16*\phm{*}&   0.0 & $\cdots$ &  0.0 &  0.0 & $\cdots$ & $>$ 0.20 \\
           &  0.21\phm{*}\phm{*}&   2.2  &  147 &  1.3 &  1.3 & $\cdots$ & \phm{$>$}       &            &  1.00*\phm{*}&   0.8  & $-$168 &  2.1 &  0.3 &  10 & \phm{$>$}       \\
J1658$+$0515 &  0.30\phm{*}\phm{*}&   0.6  & $-$104 &  0.3 &  0.3 & $\cdots$ & \phm{$>$} 0.19  & J2136$+$0041 &  2.81\phm{*}\phm{*}&   0.1  &  170 &  0.5 &  0.1 &  30 & \phm{$>$} 2.46  \\
           &  0.45\phm{*}\phm{*}&   0.0 & $\cdots$ &  0.2 &  0.2 & $\cdots$ & \phm{$>$}       &            &  3.16\phm{*}\phm{*}&   2.2  &  $-$91 &  1.1 &  0.0 &  12 & \phm{$>$}       \\
J1658$+$0741 &  0.33\phm{*}\phm{*}&   0.5  &  162 &  0.2 &  0.2 & $\cdots$ & \phm{$>$} 0.32  &            &  2.00\phm{*}\phm{*}&   1.6  & $-$113 &  0.9 &  0.9 & $\cdots$ & \phm{$>$}       \\
           &  0.96\phm{*}\phm{*}&   0.1  &  $-$11 &  0.3 &  0.3 & $\cdots$ & \phm{$>$}       &            &  0.71\phm{*}\phm{*}&   0.8  & $-$159 &  0.5 &  0.5 & $\cdots$ & \phm{$>$}       \\
           &  0.21\phm{*}\phm{*}&   7.6  &  $-$44 &  0.7 &  0.7 & $\cdots$ & \phm{$>$}       & J2158$-$1501 &  0.46\phm{*}\phm{*}&   0.1  &  $-$22 &  0.0 &  0.0 & $\cdots$ & $>$ 0.24 \\
J1723$-$6500 &  1.59\phm{*}\phm{*}&   6.6  &  134 &  1.4 &  1.0 &   9 & \phm{$>$}       &            &  0.74\phm{*}\phm{*}&   0.5  & $-$150 &  1.0 &  0.6 &   2 & \phm{$>$}       \\
           &  1.39\phm{*}\phm{*}&   0.0 & $\cdots$ &  1.1 &  0.7 & $-$12 & \phm{$>$}       & J2202$+$4216 &  0.92\phm{*}\phm{*}&   0.1  &   18 &  0.1 &  0.1 & $\cdots$ & \phm{$>$} 3.09  \\
J1733$-$1304 &  5.14\phm{*}\phm{*}&   0.0 & $\cdots$ &  0.5 &  0.3 &  20 & \phm{$>$} 1.72  &            &  1.71\phm{*}\phm{*}&   1.4  & $-$163 &  0.6 &  0.6 & $\cdots$ & \phm{$>$}       \\
J1740$+$5211 &  2.25\phm{*}\phm{*}&   0.0 & $\cdots$ &  0.4 &  0.3 & $-$28 & \phm{$>$} 1.21  &            &  1.02\phm{*}\phm{*}&   0.2  & $-$162 &  0.2 &  0.2 & $\cdots$ & \phm{$>$}       \\
J1751$+$0939 &  1.56\phm{*}\phm{*}&   0.1  &  $-$22 &  0.1 &  0.1 & $\cdots$ & \phm{$>$} 6.67  &            &  0.76\phm{*}\phm{*}&   2.8  & $-$169 &  0.8 &  0.8 & $\cdots$ & \phm{$>$}       \\
           &  0.50\phm{*}\phm{*}&   0.6  &   45 &  2.1 &  0.7 &  $-$3 & \phm{$>$}       & J2212$+$2355 &  1.00\phm{*}\phm{*}&   0.0 & $\cdots$ &  0.3 &  0.1 & $-$57 & \phm{$>$} 1.43  \\
J1800$+$7828 &  1.42\phm{*}\phm{*}&   0.0 & $\cdots$ &  0.2 &  0.2 & $\cdots$ & \phm{$>$} 2.68  & J2225$-$0457 &  3.88\phm{*}\phm{*}&   0.1  &  $-$12 &  0.4 &  0.2 &  49 & \phm{$>$} 2.31  \\
           &  0.53\phm{*}\phm{*}&   1.4  &  $-$97 &  0.8 &  0.5 &  61 & \phm{$>$}       & J2229$-$0832 &  0.43\phm{*}\phm{*}&   0.1  &   19 &  0.2 &  0.2 & $\cdots$ & \phm{$>$} 0.38  \\
           &  0.34\phm{*}\phm{*}&   0.5  &  $-$89 &  0.4 &  0.4 & $\cdots$ & \phm{$>$}       & J2253$+$1608 &  7.73\phm{*}\phm{*}&   0.0 & $\cdots$ &  0.4 &  0.4 & $\cdots$ & \phm{$>$} 2.29  \\
J1806$+$6949 &  0.58\phm{*}\phm{*}&   0.0 & $\cdots$ &  0.4 &  0.1 &  71 & \phm{$>$} 0.76  &            &  1.44\phm{*}\phm{*}&   5.2  &  $-$83 &  0.7 &  0.7 & $\cdots$ & \phm{$>$}       \\
           &  0.26\phm{*}\phm{*}&   0.7  & $-$102 &  0.9 &  0.2 & $-$78 & \phm{$>$}       &            &  0.90\phm{*}\phm{*}&   7.0  &  $-$72 &  1.4 &  1.4 & $\cdots$ & \phm{$>$}       \\
J1824$+$5651 &  0.91\phm{*}\phm{*}&   0.1  & $-$168 &  0.6 &  0.1 &  17 & \phm{$>$} 0.99  & J2320$+$0513 &  0.21\phm{*}\phm{*}&   0.5  &  147 &  0.3 &  0.2 &  42 & \phm{$>$} 0.20  \\
           &  0.27\phm{*}\phm{*}&   1.0  & $-$163 &  0.9 &  0.2 &  46 & \phm{$>$}       &            &  0.54\phm{*}\phm{*}&   0.1  &  $-$16 &  0.6 &  0.3 & $-$12 & \phm{$>$}       \\
           &  0.11\phm{*}\phm{*}&   1.8  & $-$157 &  0.4 &  0.4 & $\cdots$ & \phm{$>$}       & J2329$-$4730 &  0.41\phm{*}\phm{*}&   0.0 & $\cdots$ &  0.0 &  0.0 & $\cdots$ & $>$ 0.30 \\
J1902$+$3159 &  0.64\phm{*}\phm{*}&   0.1  &   42 &  0.4 &  0.2 & $-$11 & \phm{$>$} 0.41  & J2331$-$1556 &  0.45\phm{*}\phm{*}&   0.0 & $\cdots$ &  0.0 &  0.0 & $\cdots$ & $>$ 0.16 \\
           &  0.23\phm{*}\phm{*}&  15.5  &  119 &  1.2 &  1.2 & $\cdots$ & \phm{$>$}       &            &  0.27\phm{*}\phm{*}&   2.8  &  109 &  1.9 &  0.0 & $-$35 & \phm{$>$}       \\
           &  0.22\phm{*}\phm{*}&   0.9  &  127 &  0.7 &  0.2 & $-$36 & \phm{$>$}       & J2348$-$1631 &  0.75\phm{*}\phm{*}&   0.2  &  171 &  0.0 &  0.0 & $\cdots$ & $>$ 0.19 \\
J1924$-$2914 &  1.97\phm{*}\phm{*}&   0.2  &  158 &  0.0 &  0.0 & $\cdots$ & $>$ 0.76 &            &  0.71\phm{*}\phm{*}&   1.9  &  124 &  0.6 &  0.6 & $\cdots$ & \phm{$>$}       \\
           &  6.42\phm{*}\phm{*}&   1.1  &    3 &  0.5 &  0.5 & $\cdots$ & \phm{$>$}       &  \\
J1939$-$1525 &  0.64\phm{*}\phm{*}&   0.0 & $\cdots$ &  0.1 &  0.1 & $\cdots$ & \phm{$>$} 1.43  &  \\
J2000$-$1748 &  2.00\phm{*}\phm{*}&   0.1  &  138 &  0.3 &  0.2 & $-$47 & \phm{$>$} 2.53  &  \\
J2005$+$7752 &  1.27\phm{*}\phm{*}&   0.0 & $\cdots$ &  0.6 &  0.1 & $-$84 & \phm{$>$} 0.77  &  \\
           &  0.17\phm{*}\phm{*}&   1.2  &  $-$99 &  0.3 &  0.3 & $\cdots$ & \phm{$>$}       &  \\
J2022$+$6136 &  0.13\phm{*}\phm{*}&   1.0  &   62 &  0.8 &  0.2 & $-$29 & \phm{$>$} 0.04  &  \\
           &  1.19\phm{*}\phm{*}&   6.8  &   33 &  0.9 &  0.7 &  58 & \phm{$>$}       &  \\
\enddata
\tablecomments{ Col. $(2)$: The integrated flux density of the component,
\lq\lq $*$" and \lq\lq $dag$" denote overlapping components (see text) \newline
Col. $(3)$: The radial distance of the component center from the
center of the map\newline
Col. $(4)$: The position angle of the center of the component,
measured counterclockwise from an imaginary vertical line from the map center
towards the north\newline
Col. $(5)$: The FWHM of the major axis of the component\newline
Col. $(6)$: The FWHM of the minor axis of the component\newline
Col. $(7)$: The position angle of the major axis of the component,
measured from the north towards the east\newline
Col. $(8)$: The brightness temperature of the component
}
\end{deluxetable}
\end{landscape}

\clearpage

\begin{figure}
\caption {The distribution of observed sources in the VSS source list:
Each experiment is denoted by its observation code, found by
cross-referencing the horizontal and vertical labels.  The source
numbering is in decreasing order of {\it total flux density} which is
shown to the right of each line. The darker squares indicate the
sources presented in this paper. The lighter shaded squares are sources
observed, but not yet fully reduced as of the end of 2002, 
or of suspect quality. Many of
these have been rescheduled. The white squares are Survey sources not
yet observed as of the end of 2002.  }
\end{figure}

\clearpage

\begin{figure}
\epsscale{0.8}
\caption {\small Images of the Survey sources: 
For each source
three separate panels are presented horizontally across the page.
The first panel shows a plot of
the $(u,v)$ coverage, with $u$ on the horizontal
axis and $v$ on the vertical axis. Both axes are measured in units
of M$\lambda$. The second panel shows a plot of
the amplitude of the
visibilities (in Janskys) versus $(u,v)$ radius,
with the latter again measured in M$\lambda$.
For both of these plots only data which were actually used to make the
final image are shown.
Finally, a contour plot of the
cleaned image is shown on the right.
The contour levels are expressed as a percentage of the peak flux
density, and have the following pattern:
0.5\%, 1\%, 2\%, 4\%, 6\%, 8\%, 10\%, 15\%, 25\%, 50\%, 99\%,
along with an additional negative contour, equal in magnitude to 
the minimum positive contour level.
The peak flux
density in mJy, minimum contour level,
and synthesized HPBW in milliarcsec are
shown on the top border.
}
\end{figure}

\clearpage

\setcounter{page}{32}

\epsscale{1}
\newpage
\newpage
\newpage
\newpage
\newpage
\newpage
\newpage
\newpage
\newpage
\newpage
\newpage
\newpage
\newpage
\newpage
\newpage
\newpage
\newpage
\newpage
\newpage
\newpage
\newpage
\newpage
\newpage
\newpage

\clearpage

\begin{figure}
\caption {The histogram of core brightness temperatures:
The core brightness temperature for the 101 Survey
sources with identified cores (i.e. excluding J1723$-$6500)
has been binned, with $\log_{10}(T_b)$
on the abscissa,
and the frequency/bin as the ordinate. The open histogram 
has been used for
those sources with resolved cores.
The shaded portion of the histogram represents 
brightness temperature lower limits for unresolved cores.
On the left is the brightness temperature
distribution in the observer's frame,
while the histogram on the right represents the 
brightness temperature distribution in the source frame.
The source frame histogram only depicts brightness temperatures
from 98 of the sources, since
the redshift for three of the 101 sources was unknown.
}
\label{ratediff}
\end{figure}

\clearpage
\begin{figure}
\caption {The histogram of core angular sizes: 
The core area for
the 101 Survey
sources with identified cores (i.e. excluding J1723$-$6500) has been binned, 
with $\log_{10}(\rm{core~area})$
on the abscissa,
and the frequency/bin as the ordinate. The open histogram
has been used for
those sources with resolved cores.
The shaded portion of the histogram represents
upper limits on the core size for unresolved cores.
}
\end{figure}


\begin{thebibliography}{}

\bibitem[Asada et al.(2000)]{asa00}
Asada, K., Kameno, S., Inoue, M., Shen, Z.-Q., Horiuchi, S., \&
Gabuzda, D. C. 
2000,
in Proceedings of the VSOP Symposium, January 2000,
Astrophysical Phenomena Revealed by Space VLBI,
ed. H. Hirabayashi, P. G. Edwards, \& D. W. Murphy
(Sagamihara: The Institute of Space and Astronautical Science),
51


\bibitem[Condon et al.(1983)]{con83}
Condon, J. J., Condon, M. A., Broderick, J. J., \& Davis, M. M.
1983,
\aj,
88, 20

\bibitem[Drinkwater et al.(1997)]{dri97}
Drinkwater, M. J., et al.
1997,
\mnras,
284, 85

\bibitem[Edwards et al.(2000)]{edw00}
Edwards, P. G., Giovannini,  G., Cotton, W. D., Feretti, L.,
Fujisawa, K., Hirabayashi, H., Lara, L., \& Venturi, T. 
2000,
\pasj,
52, 1015

\bibitem[Edwards et al.(2002)]{edw02}
Edwards, P. G., Hirabayashi, H., Fomalont, E. B., Gurvits, L. I.,
Horiuchi, S., Lovell, J. E. J., Moellenbrock, G. A.,
\& Scott, W. K.
2002,
in ASP Conf. Ser., 289,
The Proceedings of the 8th IAU Asian-Pacific Meeting, Volume II,
ed. S. Ikeuchi, J. Hearnshaw, \& T. Hanawa
(San Francisco, ASP),
375

\bibitem[Fey, Clegg, \& Fomalont(1996)]{fey96}
Fey, A. L., Clegg, A. W., \& Fomalont, E. B.
1996,
\apjs,
105, 299

\bibitem[Fey \& Charlot(1997)]{fey97}
Fey, A. L., \& Charlot, P.
1997,
\apjs,
111, 95

\bibitem[Fey \& Charlot(2000)]{fey00}
Fey, A. L., \& Charlot, P.
2000,
\apjs,
128, 17

\bibitem[Fomalont et al.(2000a)]{fom00a}
Fomalont, E., et al.
2000a,
in Proceedings of the VSOP Symposium, January 2000,
Astrophysical Phenomena Revealed by Space VLBI,
ed. H. Hirabayashi, P. G. Edwards, \& D. W. Murphy
(Sagamihara: The Institute of Space and Astronautical Science),
167

\bibitem[Fomalont et al.(2000b)]{fom00b}
Fomalont, E. B., Frey, S., Paragi, Z., Gurvits, L. I., Scott, W. K.,
Taylor, A. R., Edwards, P. G., \& Hirabayashi, H.
2000b,
\apjs,
131, 95

\bibitem[Frey et al.(1997)]{fre97}
Frey, S., Gurvits, L. I., Kellermann, K. I., Schilizzi, R. T., \&
Pauliny-Toth, I.I.K. 
1997, 
\aap,
325, 511

\bibitem[Frey et al.(2000)]{fre00}
Frey, S., et al.
2000,
\pasj,
52, 975

\bibitem[Gabuzda, Pushkarev,  \& Cawthorne(1999)]{gab99}
Gabuzda, D. C., Pushkarev, A. B., \& Cawthorne, T. V.
2001,
MNRAS,
307, 725

\bibitem[Gabuzda \& G\'{o}mez(2001)]{gab01}
Gabuzda, D. C., \& G\'{o}mez, J. L.
2001,
MNRAS,
320, L49

\bibitem[Gregory et al.(1996)]{gre96}
Gregory, P. C., Scott, W. K., Douglas, K., \& Condon, J. J.  
1996,
\apjs,
103, 427

AIPS Reference
\bibitem[Greisen(1988)]{gre88}
Greisen, E. W.
1988,
in Acquisition, Processing and Archiving of Astronomical Images, 
ed. G. Longo \& G. Sedmak
(Napoli: Osservatorio Astronomico di Capodimonte), 
125

\bibitem[Griffith \& Wright(1993)]{gri93}
Griffith, M. R., \& Wright, A. E. 
1993,
\aj,
105, 1066


\bibitem[Hewitt \& Burbidge(1987)]{hew87}
Hewitt, A. \& Burbidge, G.
1987, 
\apjs, 
63, 1

\bibitem[Hewitt \& Burbidge(1993)]{hew93}
Hewitt, A. \& Burbidge, G.
1993, 
\apjs, 
87, 451

\bibitem[Hirabayashi et al.(1998)]{hir98}
Hirabayashi, H., et al. 
1998, 
Science, 
281, 1825

\bibitem[Hirabayashi et al.(2000a)]{hir00a}
Hirabayashi, H., et al.  
2000a,
\pasj,
52, 955

\bibitem[Hirabayashi et al.(2000b)]{hir00b}
Hirabayashi, H., et al.  
2000b,
\pasj,
52, 997 (Paper I)


\bibitem[Horiuchi et al.(2004)]{hor04}
Horiuchi, S., et al.
2004, 
\apj, submitted (Paper IV)

\bibitem[Jackson et al.(2002)]{jac02}
Jackson, C. A., Wall, J. V., Shaver, P. A., Kellermann, K. I.,
Hook, I. M., \& Hawkins, M. R. S.
2002,
\aap,
386, 97


\bibitem[Junor et al.(2000)]{jun00}
Junor, W., Biretta, J. A., Owen, F. N., \& Begelman, M. C.
2000,
in Proceedings of the VSOP Symposium, January 2000,
Astrophysical Phenomena Revealed by Space VLBI,
ed. H. Hirabayashi, P. G. Edwards, \& D. W. Murphy
(Sagamihara: The Institute of Space and Astronautical Science),
13


\bibitem[Kellermann \& Pauliny-Toth(1969)]{kel69}
Kellermann, K. I. \& Pauliny-Toth, I. I. K.
1969,
\apj,
155, L71

\bibitem[Kellermann et al.(1998)]{kel98}
Kellermann, K. I., Vermeulen, R. C.,
Zensus, J. A., \& Cohen, M. H.
1998,
\aj,
115, 1295


\bibitem[Lara et al.(1997)]{lar97}
Lara, L., Muxlow, T. W. B., Alberdi, A., Marcaide, J. M.,
Junor, W., \& Saikia, D. J.
1997,
\aap,
319, 405

\bibitem[Lawrence et al.(1986)]{law86}
Lawrence, C. R., Bennet, C. L., Hewitt, J. N., Langston, G. I.,
Klotz, S. E., Burke, B. F., \& Turner, K. C. 
1986,
\apjs,
61, 105

\bibitem[Lawrence et al.(1996)]{law96}
Lawrence, C. R., Zucker, J. R., Readhead, C. S., Unwin, S. C.,
Pearson, T. J., \& Xu, W.
1996,
\apjs,
107, 541


\bibitem[Levy et al.(1989)]{lev89}
Levy, G. S., et al.
1989,
\apj,
336, 1098

\bibitem[Linfield et al.(1989)]{lin89}
Linfield, R. P., et al.
1989,
\apj,
336, 1105

\bibitem[Linfield et al.(1990)]{lin90}
Linfield, R. P., et al.
1990,
\apj,
358, 350

\bibitem[Lister et al.(2001)]{lis01}
Lister, M. L., Tingay, S. J., Murphy, D. W., Piner, B. G.,
Jones, D. L., \& Preston, R. A.
2001,
\apj,
554, 948


\bibitem[Lobanov \& Zensus(2001)]{lob01}
Lobanov, A. P. \& Zensus, J. A.
2001,
Science,
294, 128

\bibitem[Lovell et al.(2000a)]{lov00a}
Lovell, J. E. J., et al.
2000a,
in Proceedings of the VSOP Symposium, January 2000,
Astrophysical Phenomena Revealed by Space VLBI,
ed. H. Hirabayashi, P. G. Edwards, \& D. W. Murphy
(Sagamihara: The Institute of Space and Astronautical Science),
183

\bibitem[Lovell(2000b)]{lov00b}
Lovell, J.
2000b,
in Proceedings of the VSOP Symposium, January 2000,
Astrophysical Phenomena Revealed by Space VLBI,
ed. H. Hirabayashi, P. G. Edwards, \& D. W. Murphy
(Sagamihara: The Institute of Space and Astronautical Science),
301

\bibitem[Lovell et al.(2004)]{lov04}
Lovell, J. E. J., et al.
2004,
\apjs, in press (Paper II)


\bibitem[Moellenbrock et al.(2000)]{moe00}
Moellenbrock, G. A., et al.
2000,
in Proceedings of the VSOP Symposium, January 2000,
Astrophysical Phenomena Revealed by Space VLBI,
ed H. Hirabayashi, P. G. Edwards, \& D. W. Murphy
(Sagamihara: The Institute of Space and Astronautical Science),
177

\bibitem[Mosoni et al.(2002)]{mos02}
Mosoni, L., Frey, S., Paragi, Z.,. Fejes, I., Edwards, P. G.,
Fomalont, E. B., Gurvits, L. I., \& Scott, W. K. 2002, 
in Proceedings of the 6th
European VLBI Network Symposium, 
ed. E. Ros, R. W. Porcas, A. P. Lobanov, \& J. A.
Zensus (Bonn: Max-Planck-Institute f\"{u}r Radioastronomie), 97


\bibitem[Okayasu et al.(2000)]{oka00}
Okayasu, R., et al.
2000,
Advances in Space Res.,
26, 681

\bibitem[Pearson \& Readhead(1988)]{pea88}
Pearson, T. J., \& Readhead, C. S.
1988,
\apj,
328, 114

\bibitem[Piner et al.(2000)]{pin00}
Piner, B. G., Edwards, P. G., Wehrle, A. E., Hirabayashi, H.,
Lovell, J. E. J., \& Unwin, S. C.
2000,
\apj,
537, 91

\bibitem[Shen et al.(1999)]{she99}
Shen, Z.-Q., Edwards, P. G., Lovell, J. E. J., 
Fujisawa, K., Kameno, S., \& Makoto, I.
1999,
\pasj,
51, 513

Difmap Reference
\bibitem[Shepherd(1997)]{she97}
Shepherd, M. C.
1997,
in ASP Conf. Ser. 125,
Astronomical Data Analysis Software and Systems VI,
ed. J. A. Zensus, G. B. Taylor, \& J. M. Wrobel
(San Francisco: ASP),
77


\bibitem[Slysh et al.(2001)]{sly01}
Slysh, V. I., et al.
2001, 
Astron. Lett., 
27, 277

\bibitem[Stanghellini et al.(1997)]{sta97}
Stanghellini, C., O'Dea, C. P., Baum, S. A.,
Fanti, R., Fanti, C.
1997,
\aap,
325,943

\bibitem[Stickel \& K\"uhr (1996)]{sti96}
Stickel, M., \& K\"uhr, H.
1996,
\aap,
115,11

\bibitem[Thompson et al.(1990)]{tho90}
Thompson, D. J., Djorgovski, S., \& De Carvalho, R.
1990,
\pasp,
102, 1235

\bibitem[Tingay et al.(1997)]{tin97}
Tingay, S. J., et al.
1997,
\aj,
113, 2025

\bibitem[Tingay et al.(2001)]{tin01}
Tingay, S. J., et al.
2001,
\apjl,
549, L55

\bibitem[Tingay et al.(2002)]{tin02}
Tingay, S. J., et al.
2002,
\apjs,
141, 311

\bibitem[Tingay et al.(2003)]{tin03}
Tingay, S. J., Jauncey, D. L., King, E. A., Tzioumis, A. K.,
Lovell, J. E. J., \&  Edwards, P. G.
2003,
\pasj,
55, 351

\bibitem[Tschager et al.(2000)]{tsc00}
Tschager, W., Schilizzi, R. T., R\"{o}ttgering, H. J. A.,
Snellen, I. A. G., \& Miley, G. K.
2000,
\aap,
360, 887

\bibitem[Tzioumis et al.(1989)]{tzi89}
Tzioumis, A. K., et al.
1989,
\aj,
98, 36


\bibitem[Vermeulen, Readhead, \& Backer (2002)]{ver94}
Vermeulen, R. C., Readhead, C. S. \& Backer, D. C.
1994,
\apj,
430, L41

\bibitem[V\'{e}ron-Cetty \& V\'{e}ron(2001)]{ver01}
V\'{e}ron-Cetty, M.-P. \& V\'{e}ron, P.
2001,
\aap,
374, 92

\bibitem[Zensus et al.(2002)]{zen02}
Zensus, J. A., Ros, E., Kellermann, K. I., Cohen, M. H.,
Vermeulen, R. C., \& Kadler, M.
2002,
\aj,
124, 662

\end{thebibliography}
\end{document}